\documentclass[reprint,prb]{revtex4-2}
\usepackage{graphicx,amssymb,amsfonts,amsmath}
\usepackage{color}
\usepackage{mathtools}
\usepackage{latexsym}
\usepackage{graphicx}
\usepackage{float}
\usepackage{dcolumn}
\usepackage{bm}
\usepackage{amssymb}
\usepackage{amsmath}
\usepackage{mathtools}
\usepackage[space]{grffile}
\usepackage{xcolor}
\usepackage{comment}
\usepackage{hyperref}
\hypersetup{
    colorlinks=true,
    linkcolor=blue,
    filecolor=blue,      
    urlcolor=blue,
    citecolor=blue
    }

\begin{document}

\title{Moment method and continued fraction expansion in Floquet Operator Krylov Space
}
\author{Hsiu-Chung Yeh}
\author{Aditi Mitra}
\affiliation{
Center for Quantum Phenomena, Department of Physics,
New York University, 726 Broadway, New York, New York, 10003, USA
}

\begin{abstract}
Recursion methods such as Krylov techniques map complex dynamics to an effective non-interacting problem in one dimension. For example,
the operator Krylov space for Floquet dynamics can be mapped to the dynamics of an edge operator of the one-dimensional Floquet inhomogeneous transverse field Ising model (ITFIM), where the latter,  after a Jordan-Wigner transformation, is a Floquet model of non-interacting Majorana fermions, and the couplings correspond to Krylov angles. We present an application of this showing that a moment method exists where given an autocorrelation function, one can construct the corresponding Krylov angles, and from that the corresponding Floquet-ITFIM. Consequently, when no solutions for the Krylov angles are obtained, it indicates that the autocorrelation is not generated by unitary dynamics. 
We highlight this by studying certain special cases: stable $m$-period dynamics derived using the method of continued fractions, exponentially decaying and power-law decaying stroboscopic dynamics. Remarkably, our examples of stable $m$-period dynamics 
correspond to $m$-period edge modes for the Floquet-ITFIM where, deep in the chain, the couplings correspond to a critical phase. Our results pave the way to engineer Floquet systems with
desired properties of edge modes and also provide examples of persistent edge modes in gapless Floquet systems.
\end{abstract}
\maketitle
\section{Introduction}

Operator dynamics takes center stage in the study of many-body quantum systems. Searching for a simple representation for operator time-evolution has been a goal for many years, and has been strongly influenced by the observation  that  ground states of many-body systems can often be efficiently simulated by recursion methods that effectively map the problem to a one-dimensional space \cite{Recbook,schollwock2005density,SCHOLLWOCK201196}. The recursion methods most commonly encountered are those where the dynamics is generated by a Hamiltonian. Here, the operator spreading can be mapped to the dynamics of a single particle on a Krylov chain with only nearest-neighbor hopping \cite{Recbook,parker2019universal,nandy2024quantum,Yates20}. The computational complexity is now encoded in the number of hopping parameters that are needed to capture the dynamics, with the Krylov chain depending both on the Hamiltonian, as well as the operator. Despite the single-particle problem having the same degree of complexity, the advantage of the mapping is that some universal features in the hopping parameters could correlate directly with some universal features of operator spreading \cite{parker2019universal,nandy2024quantum}. 

However, Hamiltonian dynamics is only a special class of unitary dynamics. In both numerical and experimental studies \cite{PAECKEL2019167998,mi2022noise,ferris2022quantum,yeh2023slowly,suchsland2023krylov}, Hamiltonian dynamics is a limit of trotterized or Floquet dynamics generated by the repeated applications of unitary gates  \cite{Suzuki1976}. In addition, Floquet dynamics have their own special features \cite{OkaRev,SondhiRev} such as a periodic spectrum reflecting the fact that energy is conserved only up to  integer multiples of a drive frequency. This can lead to new phenomena such as topological phases of matter with no counterpart in Hamiltonian systems \cite{jiang2011majorana,Rudner13,Sen13,asboth2014chiral,Khemani16,Else16,Roy17,Roy17b, Roy17c,KeyserlingkFloquet,Sondhi16a,Sondhi16b,Po16,Potter16,Potter17,Yates19,Yates20a,yeh2024productmode}.  

Recently it was shown that the Krylov space for operator spreading under a Floquet unitary can be mapped to the dynamics of an edge mode of a single particle Floquet unitary on a one-dimensional (1D) chain \cite{yeh2023universal}, the latter corresponding to the 1D inhomogeneous transverse field Ising model (ITFIM), whose couplings are given by Krylov angles. The Floquet-ITFIM has a unique topological phase diagram, and this can offer some guidelines for how an operator spreads as the operator is now interpreted as the edge mode of a trivial or a non-trivial topological phase. 

In this paper we explore further the consequences of the mapping of operator spreading to a Floquet ITFIM. We prove the existence of a moment method where given an autocorrelation function in stroboscopic time, we can construct the Krylov angles, and hence the Floquet ITFIM. We discuss the generalization of this result to the high frequency limit i.e., the limit of Hamiltonian dynamics. We also derive a continued fraction expression for the discrete Laplace transform of an autocorrelation function. The continued fraction expression is then used to derive exact expressions for Krylov angles for some simple autocorrelation functions corresponding to persistent $m$-period dynamics. In doing so, remarkably, we find a family of single-particle Floquet unitaries that are asymptotically gapless in the  bulk, i.e, the Krylov angles for the transverse field and the Ising couplings equal each other in the bulk. Nevertheless these models host $m$-period edge modes. 

The paper is organized as follows. In Section \ref{Sec: Model} we review the mapping of the Floquet operator Krylov space to the 1D Floquet-ITFIM. We then prove our assertion of a moment method, i.e., given an autocorrelation function, the Krylov angles can be obtained from them. In the process, we rule out unphysical autocorrelation functions that are impossible under unitary dynamics. We also discuss the connection to the high frequency limit, relating the Krylov angles to the more familiar Krylov hopping parameters of continuous time dynamics generated by Hamiltonians. Following this, we derive a continued fraction expression for the discrete Laplace transform of the autocorrelation function in terms of the Krylov angles. In Section \ref{Sec: Results1} we present applications of the continued fraction expression by deriving analytic expressions for the couplings of the Floquet ITFIMs that  host edge modes with persistent $m$ period dynamics for $m=2,3,4 ,6$. Each of these examples correspond to localized edge modes where the bulk of the Krylov
chain is critical. In Section \ref{sec:Results2} we present numerical results for the Krylov angles for correlators that decay with stroboscopic time in two different ways: 
as a power-law and as an exponential. We present our conclusions in 
Section \ref{Sec: Conc}. 
Five appendices provide intermediate steps in the derivation of the analytic results.

\section{Floquet Krylov chain} \label{Sec: Model}
In this section we first summarize the results of Ref.~\cite{yeh2023universal} that showed that operator spreading of any Hermitian operator under strobsocopic dynamics due to any Floquet unitary can be reproduced by a certain single-particle Floquet problem. The latter is the Floquet-ITFIM with the seed operator being its edge operator. 

A physical object on which this mapping has consequences, is the the Floquet infinite temperature autocorrelation function defined as 
\begin{align}
    A(n) = \frac{1}{\mathcal{D}}\text{Tr}[O(n)O],
\end{align}
where $\mathcal{D}$ is the Hilbert space dimension and $O$ is a Hermitian operator. $O(n)$ is defined as $O(n) = (U_F^\dagger)^n O (U_F)^n$, where $U_F$ is the Floquet unitary and $n$ is the stroboscopic time. Moreover, the Hermitian operator is normalized as follows, $\text{Tr}[O^2]/\mathcal{D} = 1$. In this definition, $|A(n)| \leq 1$.
A consequence of the mapping is that the exact same autocorrelation $A(n)$ can be reproduced by the 1D Floquet ITFIM \cite{yeh2023universal} with $O$ now corresponding to an edge operator of the the ITFIM.

The 1D Floquet ITFIM in terms of Krylov angles, and with open boundary conditions is
\begin{subequations}\label{ITFIM}
\begin{align}
&U_{\text{ITFIM}} = U_z U_{xx};\\
&U_z = \prod_{l=1}^{N/2} e^{-i\frac{\theta_{2l-1}}{2}\sigma_l^z} ;
\ U_{xx} = \prod_{l=1}^{(N-2)/2} e^{-i\frac{\theta_{2l}}{2}\sigma_l^x \sigma_{l+1}^x}.
\end{align}  
\end{subequations}
Above, we assume $N$ is even  and $\{ \theta_1, \ldots, \theta_{N-1} \} \in [0,\pi]$ are the Krylov angles. The odd Krylov angles $\theta_{2l-1}$ denote the strength of the transverse field on site $l$ while the even Krylov angles $\theta_{2l}$ denote the strength of the Ising interaction between the spins on sites $l,l+1$. 
The above model is effectively a non-interacting model. To see this, we define the Jordan-Wigner transformation
\begin{align}
  &\gamma_{2\ell -1} = \prod_{j=1}^{\ell-1}\sigma^z_j \sigma^x_\ell; &\gamma_{2\ell} = \prod_{j=1}^{\ell-1}\sigma^z_j \sigma^y_\ell.
\end{align}
The above unitary is bilinear in terms of Majorana fermions because $-i\gamma_{2l-1}\gamma_{2l} = \sigma^z_l$ and $-i\gamma_{2l}\gamma_{2l+1} =\sigma^x_l\sigma^x_{l+1} $. One can show that the Majoranas evolve under unitary evolution as follows: Under $U_z$,
\begin{subequations}
\begin{align}
    &U_z^\dagger \gamma_{2l-1} U_z = \cos(\theta_{2l-1})\gamma_{2l-1} - \sin(\theta_{2l-1})\gamma_{2l},\\
    &U_z^\dagger \gamma_{2l} U_z = \sin(\theta_{2l-1})\gamma_{2l-1} + \cos(\theta_{2l-1})\gamma_{2l}.
\end{align}
\end{subequations}
Under $U_{xx}$,
\begin{subequations}
\begin{align}
    &U_{xx}^\dagger \gamma_1 U_{xx} = \gamma_1,\quad\quad\quad 
    U_{xx}^\dagger \gamma_N U_{xx} = \gamma_N;\\
    &U_{xx}^\dagger \gamma_{2l} U_{xx} = \cos(\theta_{2l})\gamma_{2l} -\sin(\theta_{2l})\gamma_{2l+1},\\ &U_{xx}^\dagger \gamma_{2l+1} U_{xx} = \sin(\theta_{2l})\gamma_{2l} + \cos(\theta_{2l})\gamma_{2l+1}. 
\end{align}
\end{subequations}

Let us consider an operator $\Psi$ that can be written as a linear combination of Majoranas, $\Psi = \sum_{k}\psi_k\gamma_k$. The coefficients can be viewed as a column vector $\vec{\psi} = (\psi_1, \psi_2, \psi_3, \ldots)^\intercal$. Under the effect of the two unitaries, the coefficients transform according to, $U_z^\dagger \Psi U_z = \sum_{k} \psi_k' \gamma_k$ with $\vec{\psi}' = K_z \vec{\psi}$  and $U_{xx}^\dagger \Psi U_{xx} = \sum_{k} \psi_k'' \gamma_k$ with $\vec{\psi}'' = K_{xx} \vec{\psi}$. For example for $N=6$
\begin{align}
&K_z = \begin{pmatrix}
         \cos\theta_1 & \sin\theta_1 & 0 & 0 & 0 & 0 \\
 -\sin\theta_1 & \cos \theta_1 & 0 & 0 & 0 & 0 \\
 0 & 0 & \cos \theta_3 & \sin \theta_3 & 0 & 0 \\
 0 & 0 & -\sin \theta_3 & \cos \theta_3 & 0 & 0 \\
 0 & 0 & 0 & 0 & \cos \theta_5 & \sin \theta_5 \\
 0 & 0 & 0 & 0 & -\sin \theta_5 & \cos \theta_5
\end{pmatrix};\\
&K_{xx}=
\begin{pmatrix}
    1 & 0 & 0 & 0 & 0 & 0 \\
 0 & \cos \theta_2 & \sin \theta_2 & 0 & 0 & 0 \\
 0 & -\sin \theta_2 & \cos \theta_2 & 0 & 0 & 0 \\
 0 & 0 & 0 & \cos \theta_4 & \sin \theta_4 & 0 \\
 0 & 0 & 0 & -\sin \theta_4 & \cos \theta_4 & 0 \\
 0 & 0 & 0 & 0 & 0 & 1
\end{pmatrix}.
\end{align}
The unitary evolution in the Majorana basis after one time step is
\begin{widetext}
\begin{align}
    K &=  K_{xx} K_z \nonumber\\
    &= \begin{pmatrix}
        \cos \theta_1 & \sin \theta_1 & 0 & 0 & 0 & 0 \\
 -\sin \theta_1 \cos \theta_2 & \cos
   \theta_1 \cos \theta_2 & \sin \theta_2
   \cos \theta_3 & \sin \theta_2 \sin \theta_3 & 0 & 0 \\
 \sin \theta_1 \sin \theta_2 & -\cos \theta_1\sin \theta_2 & \cos \theta_2 \cos
   \theta_3 & \cos \theta_2\sin \theta_3  & 0
   & 0 \\
 0 & 0 & -\sin \theta_3 \cos \theta_4 & \cos
   \theta_3 \cos \theta_4 & \sin \theta_4
   \cos \theta_5 & \sin \theta_4 \sin \theta_5 \\
 0 & 0 & \sin \theta_3 \sin \theta_4 & -\cos \theta_3\sin \theta_4  & \cos \theta_4
   \cos \theta_5 & \cos \theta_4 \sin \theta_5  \\
 0 & 0 & 0 & 0 & -\sin \theta_5 & \cos \theta_5
\end{pmatrix}.
\label{Eq: K matrix}
\end{align}
\end{widetext}
Now we will construct the Krylov operator space for the edge operator $\gamma_1$. 
First, we consider the vector $|\gamma_1) = |1) = (1, 0, 0, \ldots)^\intercal$. The operator Krylov space is therefore spanned by $\{|1), K|1), K^2|1), \ldots \}$. By performing  the Gram-Schmidt procedure, one can represent the unitary evolution in the Krylov orthonormal basis labeled as $\{ |\mathcal{O}_0), |\mathcal{O}_1), \ldots, |\mathcal{O}_{N-1}) \}$, in the following upper-Hessenberg matrix \cite{Yates21,suchsland2023krylov,yeh2023universal}, whose explicit form for $N=6$ is
\begin{align}
    \Tilde{K}=\begin{pmatrix}
        a_0 & c_1 & c_2 & c_3 & c_4 & c_5\\
        b_1 & a_1 & \frac{a_1}{c_1}c_2 & \frac{a_1}{c_1}c_3 & \frac{a_1}{c_1}c_4 & \frac{a_1}{c_1}c_5\\
        0 & b_2 & a_2 & \frac{a_2}{c_2}c_3 & \frac{a_2}{c_2}c_4 & \frac{a_2}{c_2}c_5\\
        0 & 0 & b_3 & a_3 & \frac{a_3}{c_3}c_4 & \frac{a_3}{c_3}c_5\\
        0 & 0 & 0 & b_4 & a_4 & \frac{a_4}{c_4}c_5\\
        0 & 0 & 0 & 0 & b_5 & a_5
    \end{pmatrix}.\label{Eq: K matrix Krylov basis K6}
\end{align}
Above,
\begin{align}
&b_j=(\mathcal{O}_j|K|\mathcal{O}_{j-1});
&\frac{a_j}{c_j}c_k=(\mathcal{O}_j|K|\mathcal{O}_k)\ \text{for}\ j \leq k,
\label{Eq: K-matrix Krylov space}
\end{align}
with $a_j, b_j, c_j$ depending on the Krylov angles as follows \cite{yeh2023universal}
\begin{subequations} \label{Eq: K-matrix Krylov space2}
\begin{align}
    &a_j = \cos\theta_j\cos\theta_{j+1};\quad\quad\quad
    b_j = \sin\theta_j;\\
    &c_j = (-1)^j\cos\theta_{j+1} \prod_{k=1}^j\sin\theta_k,
    \end{align}
\end{subequations}
with $\theta_0 = \theta_{N} = 0$ and $\theta_i \in [0,\pi]$. 

Note that the $K$-matrix \eqref{Eq: K matrix} and the $\Tilde{K}$-matrix \eqref{Eq: K matrix Krylov basis K6} describe the same unitary evolution of operators but in different representation bases: the former is in the Majorana basis and the latter is in the Krylov orthonormal basis. Ref.~\cite{yeh2023universal} showed that the matrix $\tilde{K}$ with
the parametrization \eqref{Eq: K-matrix Krylov space},\eqref{Eq: K-matrix Krylov space2} can reproduce the Krylov space structure for the unitary evolution of any Hermitian operator for any Floquet system. Thus matching the Krylov angles between the ITFIM and the problem at hand provides an exact mapping. Interestingly, the cosine of the Krylov angles $\cos\theta_j$  are related to Verblunsky coefficients appearing in the CMV basis representation of unitary matrices, and derived using the method of orthogonal polynomials on the unit circle \cite{CANTERO200329,kolganov2025streamlinedkrylovconstructionclassification}.

Above we have discussed how operator spreading under Floquet dynamics is captured by a  particular single-particle Floquet problem, the ITFIM in 1D. However, there are other mappings to  single particle Floquet systems in 1D, and we discuss the relation between those to the ITFIM.
We could have alternately mapped the problem to the inhomogeneous XY model, which is also a non-interacting system. However, this model is essentially the same as the ITFIM because the XY model is two copies of transverse-field Ising model. To see this note that, $\sigma^x_l\sigma^x_{l+1} = - i\gamma_{2l}\gamma_{2l+1}$ and $\sigma^y_l\sigma^y_{l+1} = i\gamma_{2l-1}\gamma_{2l+2}$. The XX terms $\sigma^x_l\sigma^x_{l+1}$  contain the 
Majorana bilinears $\gamma_2\gamma_3, \gamma_4\gamma_5, \gamma_6\gamma_7, \ldots$, while the  YY terms $\sigma^y_l\sigma^y_{l+1}$ contain the Majorana bilinears, $\gamma_1\gamma_4, \gamma_3\gamma_6, \gamma_5\gamma_8, \ldots$. The system thus separates into two independent Kitaev chains: The first chain contains the Majoranas, $\gamma_1, \gamma_4, \gamma_5, \gamma_8, \gamma_9, \gamma_{12}, \ldots$, and the second chain contains the Majoranas, $\gamma_2, \gamma_3, \gamma_6, \gamma_7, \gamma_{10}, \gamma_{11}, \ldots$. Adding a transverse field $\sigma^z_l = -i\gamma_{2l-1}\gamma_{2l}$ also results in a free system, but one where the two Majorana chains are now coupled to each other.

We also mention a special case corresponding to all the Krylov angles being equal to $\theta_i=\pi/2$. For this case, all the elements of $\tilde{K}$ vanish except the elements of the first sub-diagonal which are $b_i=1$. This is an example of a maximally ergodic system \cite{suchsland2023krylov} where an operator spreads with no memory of its initial conditions, with the autocorrelation function
vanishing after the first time-step $A(n)= \delta_{n,0}$. Consequently, in many examples where an autocorrelation decays to zero at long times, the Krylov angles approach $\pi/2$ in the bulk of the chain \cite{yeh2023universal}. 

\subsection{Exact Moment Method}\label{Sec: algorithm}

In this section we prove a moment method where given an autocorrelation function $A(n)$, one can solve for the Krylov angles, and therefore construct the Floquet-ITFIM \eqref{ITFIM} and equivalently the full Krylov matrix \eqref{Eq: K matrix Krylov basis K6},\eqref{Eq: K-matrix Krylov space}. In addition, when the method yields no solutions of the Krylov angles,
it indicates that the given $A(n)$ is not allowed under unitary dynamics.
 
The autocorrelation is reproduced by the time evolution of the vector $|1) = (1, 0, 0, \ldots)^\intercal$ such that $A(n) = (1|K^n|1)$. Note that $(1|K^n|1) = (1|\Tilde{K}^n|1)$, and we work with $K$ here since $K$ is sparse. For $n=1$ and $2$, and using \eqref{Eq: K matrix} we have
\begin{subequations}
\begin{align}
    &A(1) = \cos\theta_1, \label{Eq: A(1)}\\
    &A(2) = \cos^2\theta_1 - \sin^2\theta_1 \times \cos\theta_2.\label{Eq: A(2)}
\end{align}
\end{subequations}
In general, $A(n)$ has the following form
\begin{align}
    A(n) = f(n-1) + (-1)^{n-1}\cos\theta_n \prod_{k=1}^{n-1} \sin^2\theta_k, \label{Eq: Autocorrelation to angle}
\end{align}
where $f(n-1)$ is some function that depends on the Krylov angles $\{\theta_1, \theta_2, \ldots, \theta_{n-1}\}$. Note that the dependence on $\theta_n$ of $A(n)$ only appears in the second term. Each term can be interpreted as the amplitude for a free particle to hop on the chain, with the matrix element $K_{jk}$ representing the hopping amplitude from site $k$ to site $j$. We define site $1$ to be the left end of the chain. The $\theta_n$ dependence comes from the following hopping trajectory: a particle starting from site $1$ travels to the right and visits site $n$ before  returning to site $1$ at time step $n$.  In each step of the time evolution under \eqref{Eq: K matrix}, the maximal hopping distance is $2$ sites. Therefore, a particle must spend the first half of the time traveling to the site $n$, returning to site $1$ during the second half of the time. According to  \eqref{Eq: K matrix}, the hopping distance is biased and depends on  even or odd sites as follows
\begin{align}
    K_{j,2l} \neq 0 \text{ and } K_{j,2l-1} \neq 0 \text{ when } 2l+1 \geq j \geq 2l-2,
    \label{Eq: Floquet hopping}
\end{align}
where $l$ is a positive integer.
\begin{figure}[h!]
    \includegraphics[width=0.4\textwidth]{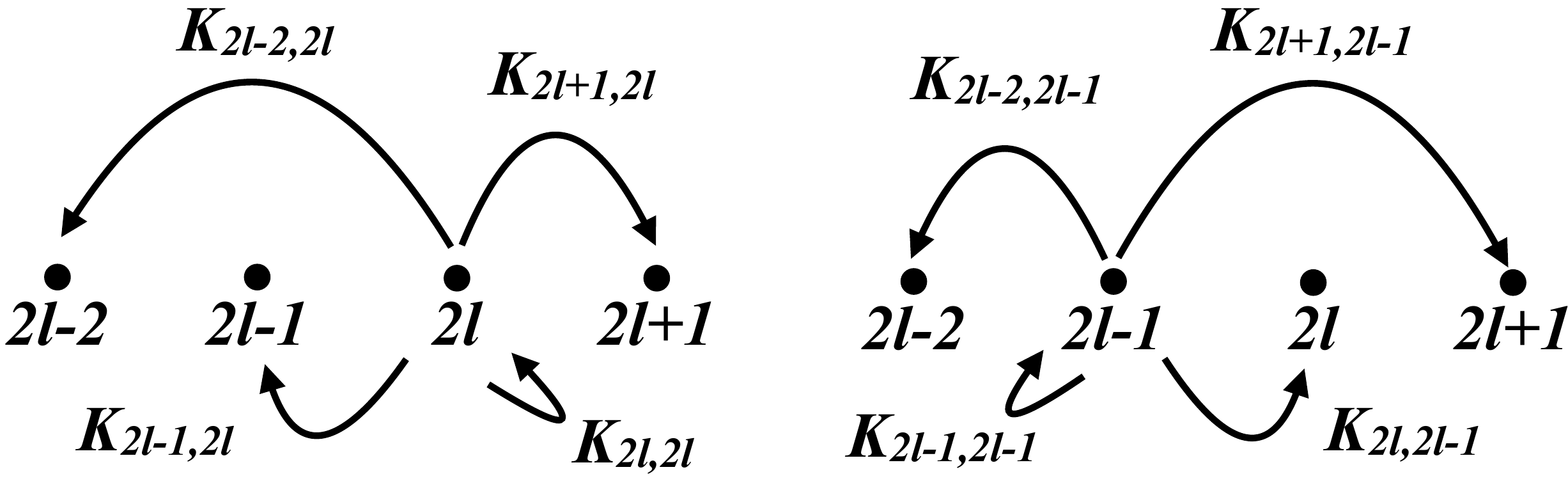}

    \caption{The pictorial illustration of \eqref{Eq: Floquet hopping}. The hopping distance is biased and depends on the location of the particle being even or odd.}
    \label{Fig: Floquet hopping}
\end{figure}

When the particle is on the even (odd) site, it can maximally hop to the right with distance 1 (2) site but to the left with distance 2 (1) sites, see
Fig.~\ref{Fig: Floquet hopping}. Therefore, for a trajectory to start from site $1$,  visit site $n$, and then return to site $1$ within $n$ time steps, the particle has to travel to the right (left) by hopping on odd (even) sites, except for the turning point. Thus when $n$ is even, $n = 2l$ with $l$ being a positive integer, the only trajectory that depends on $\theta_n$ is 
\begin{align}
    &K_{12}K_{24}\ldots K_{2l-2,2l} K_{2l,2l-1} K_{2l-1,2l-3} \ldots K_{53}K_{31}\nonumber\\
    =& K_{12} \left(\prod_{k=2}^{l} K_{2k-2,2k}\right)K_{2l,2l-1}\left(\prod_{k=1}^{l-1}K_{2k+1,2k-1} \right)\nonumber\\
    =& \sin\theta_1 \left(\prod_{k=2}^{l} \sin\theta_{2k-2}\sin\theta_{2k-1} \right)(-\sin\theta_{2l-1}\cos\theta_{2l})\nonumber\\
    &\times\left(\prod_{k=1}^{l-1}\sin\theta_{2k-1}\sin\theta_{2k} \right)\nonumber\\
    =& -\cos\theta_{n} \prod_{k=1}^{n-1} \sin^2\theta_{k}, \text{ with } n=2l,
\end{align}
where we use that $K_{12} = \sin\theta_1$, $K_{2k-2,2k} = \sin\theta_{2k-2}\sin\theta_{2k-1}$, $K_{2l,2l-1} = -\sin\theta_{2l-1}\cos\theta_{2l}$ and $K_{2k+1,2k-1} = \sin\theta_{2k-1}\sin\theta_{2k}$. The above result agrees with the second term in \eqref{Eq: Autocorrelation to angle} when $n$ is even.

Similarly, when $n$ is odd, $n = 2l-1$, the only trajectory that depends on $\theta_n$ is
\begin{align}
    &K_{12}K_{24}\ldots K_{2l-4,2l-2} K_{2l-2,2l-1} K_{2l-1,2l-3} \ldots K_{53}K_{31}\nonumber\\
    =& K_{12} \left(\prod_{k=2}^{l-1} K_{2k-2,2k}\right)K_{2l-2,2l-1}\left(\prod_{k=1}^{l-1}K_{2k+1,2k-1} \right)\nonumber\\
    =& \sin\theta_1 \left(\prod_{k=2}^{l-1} \sin\theta_{2k-2}\sin\theta_{2k-1} \right)(\sin\theta_{2l-2}\cos\theta_{2l-1})\nonumber\\
    &\times\left(\prod_{k=1}^{l-1}\sin\theta_{2k-1}\sin\theta_{2k} \right)\nonumber\\
    =& \cos\theta_{n} \prod_{k=1}^{n-1} \sin^2\theta_{k}, \text{ with } n=2l-1,
\end{align}
where we use that $K_{2l-2,2l-1} = \sin\theta_{2l-2}\cos\theta_{2l-1}$. This also agrees with \eqref{Eq: Autocorrelation to angle} when $n$ is odd. Therefore, \eqref{Eq: Autocorrelation to angle} is valid for general $n$.

One can solve the Krylov angles order by order from a given dataset of $A(n)$ such that the Gram-Schmidt procedure in operator space can be bypassed. Since the Krylov angle is defined within $[0,\pi]$, one obtains an unique solution for $\cos\theta_n$ at each order $n$ and hence an unique solution of $\theta_n$. As discussed above, the second term in \eqref{Eq: Autocorrelation to angle} comes from a particular trajectory: the return amplitude for a particle traveling from the first to $n$-th site and back to the first site at $t = n$, under unitary time evolution. The first term in \eqref{Eq: Autocorrelation to angle} corresponds to the remaining trajectories contributing to the return amplitude, and in particular those that never reach site $n$. 

Numerically, the Krylov angles are solved order by order according to
\begin{align}
    \cos\theta_n = \frac{A(n)-f(n-1)}{(-1)^{n-1}\prod_{k=1}^{n-1} \sin^2\theta_k}.\label{algo2}
\end{align}
One assumes the $K$ matrix \eqref{Eq: K matrix} with all angles initially unknown.
The numerical algorithm involves using the relation $A(n) = (1|K^n|1)$ where $A(n)$ is
obtained via numerical simulations such as exact diagonalization (ED), or as in the examples below, we just assume some analytic form for it. Then,
at each stroboscopic time, equating the left and right sides determine $\theta_n$. When
determining $\theta_{n}$, the actual numerical values of $\theta_1\ldots \theta_{n-1}$ are entered. 
Thus, the evaluation $(1|K^n|1)$ gives us $f_{n-1} + c_{n-1}\cos\theta_{n}$ where
$f_{n-1},c_{n-1}$ are numbers. Equating this to $A(n)$ determines $\cos\theta_{n}$. This numerical algorithm might fail when $\prod_{k=1}^{n-1} \sin^2\theta_k$ becomes too small, reaching numerical precision. The numerical precision depends on the precision of $A(n)$ and the cumulative numerical error in solving for $\theta_n$. If $\theta_n$ approaches $\pi/2$ fast enough, which corresponds to the correlator decaying to zero fast enough, $\prod_{k=1}^{n-1} \sin^2\theta_k$ will saturate to a large enough constant value such that the numerical algorithm is stable.

\begin{figure}[h!]
    \includegraphics[width=0.4\textwidth]{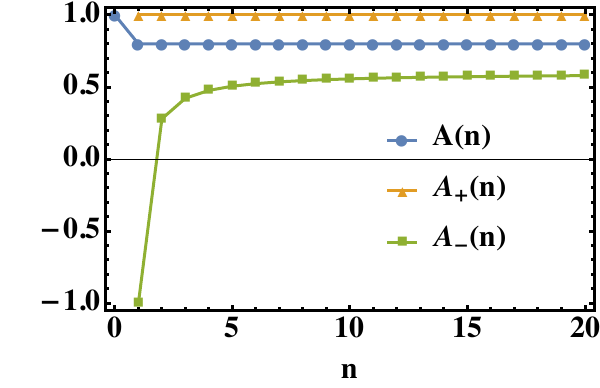}

    \caption{Illustration of the constraint on an autocorrelation obtained from unitary evolution. We set $A(n\geq 1) = 0.8$ and the upper/lower bound $A_+(n)/A_-(n)$ is numerically computed from \eqref{Eq: A constraint}. As $n$ increases, the range of the bound goes from $[-1,1]$ to about $[0.6, 1]$ for this example.}
    \label{Fig: Unitary Constraint}
\end{figure}

\eqref{algo2} also provides constraints on $A(n)$. In particular, if $A(n)$ is generated by unitary evolution, and since $|\cos\theta|\leq 1$,  $A(n)$ must satisfy the following condition
\begin{subequations}
\begin{align}
    &A_{-}(n) \leq A(n) \leq A_{+}(n),\\ &A_{\pm}(n) = f(n-1) \pm \prod_{k=1}^{n-1} \sin^2\theta_k.
    \label{Eq: A constraint}
\end{align}
\end{subequations}
Note that $A_\pm(n)$ can be determined from $\{ A(1), A(2), \ldots, A(n-1)\}$, i.e., all the previous $n-1$ Krylov angles. The past information of $A(n)$ will constrain the range of $A(n)$ that is allowed by unitary dynamics. In Fig.~\ref{Fig: Unitary Constraint}, we numerically computed the example of $A(n\geq 1) = 0.8$, where $A(0)=1$ always. At $n = 1$, the unitary constraint allows $A(1)$ to be within $[-1,1]$. By choosing $A(1) = 0.8$, it immediately shrinks the range of $A(2)$ to about $[0.3,1]$. If one has an autocorrelation $A(1\leq n \leq n^*) = 0.8$ for some large $n^*$, the unitary evolution only allows $A(n^*+1)$ within the range of about $[0.6,1]$ for this example. This highlights that the algorithm \eqref{algo2} will fail if the input data of the autocorrelation is non-unitary.
The algorithm also shows that there is a direct relation between the number of Krylov angles and the stroboscopic time upto which the dynamics is known, and vice versa. 

A natural question that arises is how does this algorithm for generating Krylov parameters from the autocorrelation function generalizes to Hamiltonian dynamics, i.e., the limit of vanishingly small stroboscopic time. Below, we briefly mention a similar algorithm in the Hamiltonian limit. In contrast to stroboscopic dynamics discussed above, time is a continuous parameter in the Hamiltonian limit. It is more convenient to construct the operator Krylov space by applying the Liouvillian superoperator $\mathcal{L} = i[H,\cdot]$ to the initial operator. The Liouvillian superoperator is the following  tridiagonal matrix in the Krylov orthonormal basis
\begin{align}
    \mathcal{L} = \begin{pmatrix}
        0 & -\Tilde{b}_1 & 0 & 0 & 0 & \ldots\\
        \Tilde{b}_1 & 0 & -\Tilde{b}_2 & 0 & 0 & \ldots\\
        0 & \Tilde{b}_2 & 0 & -\Tilde{b}_3 & 0 & \ldots\\
        0 & 0 & \Tilde{b}_3 & 0 & -\Tilde{b}_4 & \ddots\\
        0 & 0 & 0 & \Tilde{b}_4 & 0 & \ddots\\
        \vdots & \vdots & \vdots & \ddots & \ddots & \ddots
    \end{pmatrix}.
    \label{Eq: Liouvillian}
\end{align}
The continuous time autocorrelation $C(t)$ is related to $\mathcal{L}$ via the relation: $C(t) =  (1|e^{\mathcal{L}t}|1)$ with $|1) = (1, 0, 0, \ldots)^\intercal$ in the Krylov orthonormal basis. In contrast to the convention in \cite{parker2019universal}, we include $i$ in the definition of the Liouvillian superoperator such that \eqref{Eq: Liouvillian} is anti-symmetric.  
Defining the Fourier transform of $C(t)$ as
\begin{align}
\Phi(\omega) = \int_{-\infty}^{\infty} dt\ e^{-i\omega t}C(t),
\end{align}
the moment $m_k$ corresponds to 
\begin{align}
 m_k = \frac{1}{2\pi}\int_{-\infty}^{\infty}d\omega\ \omega^k \Phi(\omega).
 \end{align}
From direct substitution of $C(t) = \int d\omega e^{i\omega t}\Phi(\omega)/(2\pi)$ it is straightforward to see that
\begin{align}
 m_k = \lim_{t \rightarrow 0 } \frac{d^k C(t)}{i^kdt^k}, 
\end{align}
implying the following relation between the Taylor expansion of the autocorrelation function and the moments $C(t) = \sum_{k=0}^\infty m_k (it)^k/(k!)$.
This expression also shows that the moments can be expressed in terms of the Liouvillian superoperator as
\begin{align}
    m_k = (-i)^k(1|\mathcal{L}^k|1),\label{mk2}
\end{align}
which resembles the relation for the Floquet autocorrelation, $A(n) = (1|K^n|1)$. If all the moments are known, one may compute all the off-diagonal elements of $\mathcal{L}$. This way of determining $\{ \Tilde{b}_k \}$ is called the moment method, which is sometimes amenable to analytic treatments \cite{Recbook,parker2019universal,dymarsky2020quantum,dymarsky2021krylov,nandy2024quantum}. Below, we only discuss the numerical algorithm similar to the Floquet case. From \eqref{mk2}, the moment $m_k$ can be realized by summing over all trajectories that start from the first site and then return to the first site after $k$ steps according to the nearest-neighbor hopping structure of $\mathcal{L}$. Note that the zero in the diagonal of $\mathcal{L}$ indicates that all odd moments are zero. This is because the particle is only allowed to hop left or right at each step, with the full trajectory requiring even number of steps. The second and fourth moments can be explicitly written as
\begin{align}
    &m_2 = \Tilde{b}_1^2,\\
    &m_4 = \Tilde{b}_1^2 \Tilde{b}_2^2 + \Tilde{b}_1^4.
\end{align}
In general, for a positive integer $l$, $m_{2l}$ is composed of two parts
\begin{align}
    m_{2l} = g(l-1) + \Tilde{b}_1^2 \Tilde{b}_2^2 \ldots \Tilde{b}_{l}^2,
\end{align}
where $g(l-1)$ is some function that depends on $\{\Tilde{b}_1, \Tilde{b}_2, \ldots, \Tilde{b}_{l-1}\}$ and the second term is the only trajectory that depends on $\Tilde{b}_l$.

Similar to the  Floquet example, the contribution of $\Tilde{b}_l$ in the second term is a consequence of a particular trajectory where the particle reaches site $l$ and returns to the first site within $2l$ steps. Since the particle can only travel one site in each step, it must keep hopping to the right in the first $l$ steps to reach site $l$, and then return to the first site in the remaining $l$ steps. Numerically, $\Tilde{b}_l$ can be solved order by order according to
\begin{align}
    \Tilde{b}_l = \sqrt{\frac{m_{2l} - g(l-1)}{\prod_{k=1}^{l-1}\Tilde{b}_k^2}},
\end{align}
where $g(l-1)$ is now just a number when $\{\Tilde{b}_1, \Tilde{b}_2, \ldots, \Tilde{b}_{l-1}\}$ have been numerically computed in the previous steps of the iteration. The above relation between moments $\{ m_k \}$ and the off-diagonal elements $\{ \Tilde{b}_k \}$ can also be realized by Dyck paths, and an analytic recurrence relation for solving for $\Tilde{b}_k$ can be derived \cite{Recbook,parker2019universal,nandy2024quantum}.
Although the above discussion seems straightforward, there is a mathematical subtlety that the Taylor expansion, $C(t) = \sum_{k=0}^\infty m_k (it)^k/(k!)$, is not necessarily convergent for arbitrary moments. However, for physical quantum systems, $\tilde{b}_k$ is conjectured to grow at most linearly in $k$ and validates the Taylor expansion, see details in \cite{parkerPhDThesis} and references therein.
The Floquet autocorrelation function plays the same role as the moments in that it directly determines the Krylov angles. Consequently, there is a direct relation between the number of Krylov angles and the strobsocopic time.
If we have knowledge of $n$ Krylov angles, we have simulated the dynamics upto $n$ stroboscopic
time steps. In contrast, there is no direct relation between say the first $n$ Krylov parameters in the high frequency limit and the time upto which the dynamics is known. One has to be careful about the convergence of the Taylor expansion in the high frequency limit.    For example, Ref.~\cite{parker2019universal} showed that the value of the moments grows exponentially in chaotic systems and the Taylor expansion of the autocorrelation has a finite convergence radius. Thus, knowing more moments does not allow one to construct the late time dynamics beyond the convergence radius. However, in the Floquet case, the first $n$ Krylov angles can exactly determine the first $n$ time steps regardless of the model.

Finally, we discuss how the Krylov angles $\{ \theta_k \}$ are related to the $\{ \Tilde{b}_k \}$ in the high frequency limit, or equivalently the limit of small Krylov angles. The $\Tilde{K}$-matrix \eqref{Eq: K matrix Krylov basis K6} with small Krylov angles can be realized as the short time expansion of $e^{\mathcal{L}t}$: $\Tilde{K} \approx 1 + \mathcal{L}t$. The small Krylov angle expansion of \eqref{Eq: K matrix Krylov basis K6} leads to
\begin{align}
    \Tilde{K} \approx \begin{pmatrix}
        1 & -\theta_1 & 0 & 0 & 0 & \ldots\\
        \theta_1 & 1 & -\theta_2 & 0 & 0 & \ldots\\
        0 & \theta_2 & 1 & -\theta_3 & 0 & \ldots\\
        0 & 0 & \theta_3 & 1 & -\theta_4 & \ddots\\
        0 & 0 & 0 & \theta_4 & 1 & \ddots\\
        \vdots & \vdots & \vdots & \ddots & \ddots & \ddots
    \end{pmatrix}
    \approx 1 + \mathcal{L}t.  
\end{align}
Therefore, one concludes that $\theta_k = \Tilde{b}_k t$. This relation only holds for $\theta_k, \Tilde{b}_k t \ll 1$. According to \eqref{Eq: A constraint}, the size of the bound of $A(n)$ is given by
\begin{align}
    A_+(n) - A_-(n) = 2\prod_{k=1}^{n-1}\sin^2\theta_k.
\end{align}
In the high frequency limit, the two closest autocorrelation data points become infinitesimally close  ($|A(n) -A(n-1)| \rightarrow 0$ as $t \rightarrow 0$), and the bound shrinks. The Floquet dynamics can be interpreted  as a deformation of the continuous time dynamics with a finite bound whose size is related to the size of the Trotter time step.

\subsection{Discrete Laplace transformation and continued fraction}
For a given infinite temperature autocorrelation function $A(n)$, we can define its discrete Laplace transformation $G_L(z)$
\begin{align}
    G_L(z) = \sum_{n=0}^{\infty} A(n)z^{-n},
    \label{Eq: G Laplace}
\end{align}
where $z$ is complex number with $|z|>1$ ensuring a convergent expression for $G_L(z)$. 
The discrete Laplace transformation $G_L$ has an equivalent representation in terms of the continued fraction
\begin{align}
    G_C(z,\theta) = \frac{\alpha_0}{\beta_0 + \frac{\alpha_1}{\beta_1 + \frac{\alpha_2}{ \ddots}}},\label{Eq: G continued fraction}
\end{align}
where
\begin{subequations}\label{eq:alpha-beta}
\begin{align}
    &\alpha_0 = z;\quad\quad \beta_0 = z - \cos\theta_1; \label{Eq: alpha beta-1}\\
    &(\alpha_{k}, \beta_{k}) = \Big\{\begin{array}{cc}
    (\sin^2\theta_k,\ z\cos\theta_{k+1} - \cos\theta_k) & \text{if}\ k\ \text{is odd};\\
     (z^2\sin^2\theta_k,\ z\cos\theta_{k} - \cos\theta_{k+1}) & \text{else}.
    \end{array}
    \label{Eq: alpha beta-2}
\end{align}
\end{subequations}
The detailed proof of the equivalence, $G_L(z) = G_C(z,\theta)$, is shown in Appendix \ref{Sec: A}.

The general behavior of the Krylov angles $\theta_k$ for decaying autocorrelation functions was numerically studied in Ref.~\cite{yeh2023universal}, where the $\theta_k$ approach $\pi/2$ for sufficiently large $k$, i.e, in the bulk of the Krylov chain. The Krylov chain with bulk angles equalling $\pi/2$ is maximally ergodic in operator Krylov space, with the dual unitary limit corresponding to all Krylov angles being $\pi/2$ \cite{suchsland2023krylov}.  We now discuss the form of the continued fraction for the dual unitary limit when all Krylov angles are $\pi/2$. 

In the dual unitary limit, the autocorrelation becomes simple: $A(n) = \delta_{n,0}$, i.e, the autocorrelation is only non-zero at $n = 0$. Due to maximal ergodicity in operator Krylov space, the initial operator evolves into another new orthonormal operator in each time step under unitary evolution and therefore the autocorrelation is zero for $n \geq 1$. The corresponding discrete Laplace transformation is also simple: $G_L(z) = 1$ according to \eqref{Eq: G Laplace}. However, the corresponding continued fraction has the following form
\begin{align}
    &G_C(z,\pi/2) = \frac{z}{z+X},
    &X = \alpha_1/\alpha_2/\alpha_3 \ldots,
\end{align}
where $\alpha_{\rm odd}=1,\alpha_{\rm even}= z^2$ and $X$ satisfies the self consistent equation, $X = 1 /(z^2/X)$. Therefore, $X$ can be $0$ or divergent. The physical choice is $X=0$ so that $G_C(z,\pi/2) = 1$ is consistent with the Laplace transformation. For decaying autocorrelations, one can approximate the bulk of the Krylov chain as being maximally ergodic, with the latter part of the continued fraction being negligible as $X = 0$. 

In this paper, we present a different regime from the dual unitary case, it is one where the Krylov angles approach $\pi/2$ rather than being exactly $\pi/2$ everywhere on the chain. Therefore, the details of all Krylov angles are important and the continued fraction cannot be truncated. We consider two broadly different classes of autocorrelations. One where 
the autocorrelations are non-decaying in time, and show persistent $m$-period oscillations (Section ~\ref{Sec: Results1}). The second are exponentially and power-law decaying in time autocorrelations (Section ~\ref{sec:Results2}). For all these examples, we construct the Krylov angles, and therefore the corresponding $U_{\rm ITFIM}$. 

\section{Analytic solutions of persistent $m$-period autocorrelation} \label{Sec: Results1}
We consider persistent $m$-period autocorrelations of the form
\begin{align}
A^{(m)}(n>0) = A\cos(2\pi n/m), \,\, A^{(m)}(n=0)=1. \label{Adef}
\end{align}
where $A$ is a real number within $[0,1]$ and $m$ is an integer. Below the analytic solution of the Krylov angles for $m = 1, 2, 3, 4, 6$  are presented. These examples give simple analytic expressions for the Krylov angles. In particular, the cosine of the Krylov angles can be formulated as a rational function of the Krylov index $k$ and with only a linear dependence on $A$ in both the numerator and the denominator. However, this is not true for generic $m$-period autocorrelations for which one can only obtain the numerical values of the Krylov angles employing the moment-method algorithm described in the previous section. 

We find that the five examples $m=1,2,3,4,6$ corresponding to the autocorrelation function \eqref{Adef} divide into two different classes. In each class, the solutions are related by the reflection, $z \rightarrow -z$ or square-reflection, $z \rightarrow -z^2$  transformation of the discrete Laplace transformation 
\begin{align}
G_L^{(m)}(z) = \sum_{n=0}^\infty A^{(m)}(n)z^{-n}.
\end{align}

\subsection{$1,2,4$ period autocorrelation}
\begin{figure*}
    \includegraphics[width=0.32\textwidth]{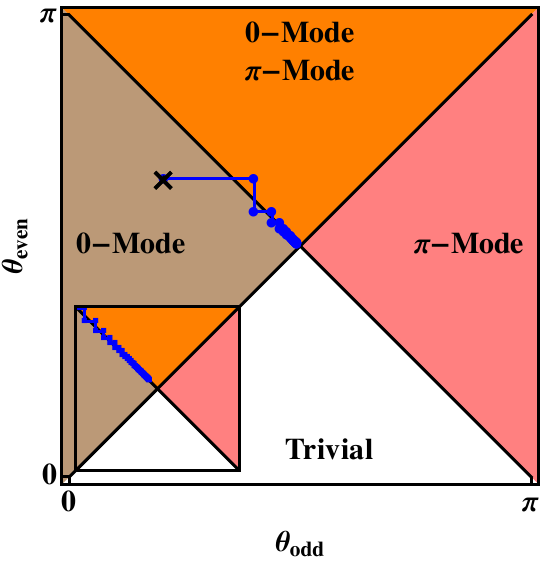}
    \includegraphics[width=0.32\textwidth]{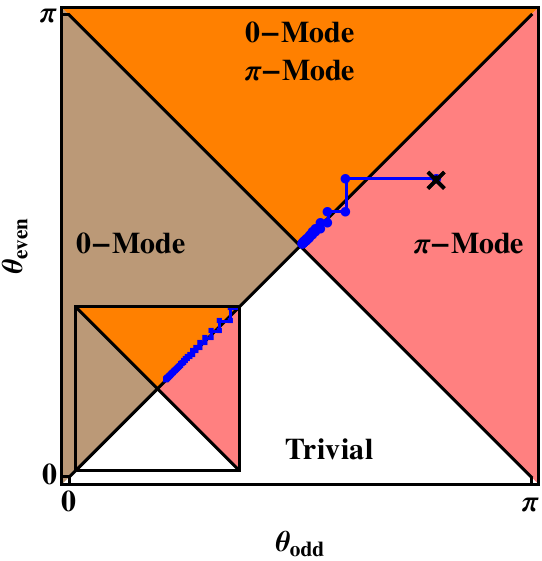}
    \includegraphics[width=0.32\textwidth]{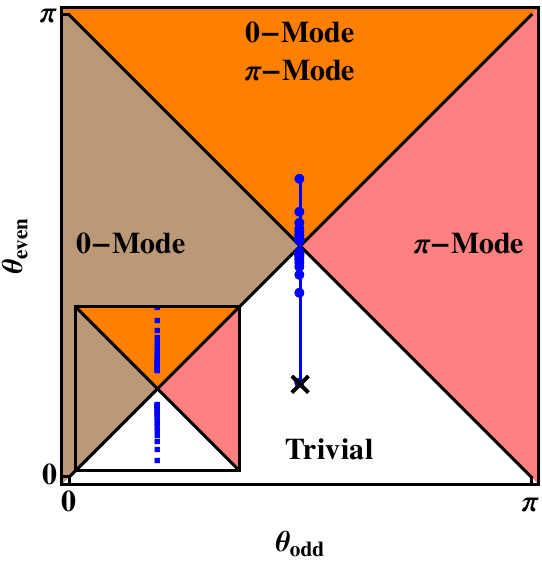}

    \caption{The trajectories of first hundred Krylov angles for \eqref{Eq: m=1 Krylov angle} ($m=1$ period, left panel), \eqref{Eq: m=2 Krylov angle} ($m=2$ period, middle panel) and \eqref{Eq: m=4 Krylov angle} ($m=4$ period, right panel) with $A = 0.8$.  The black crosses
    denote the starting point  $(\theta_1,\theta_2)$ of the trajectory. The four shaded triangular regions mark the four homogeneous topological phases of the ITFIM $\theta_1 = \theta_3 = \theta_5 = \ldots = \theta_{\text{odd}}$ and  $\theta_2 = \theta_4 = \theta_6 = \ldots = \theta_{\text{even}}$, with the phases labeled by the nature of the stable edge modes they support. All trajectories approach the center of the phase diagram where the bulk of the ITFIM is critical $\theta_{\text{even}}=\theta_{\text{odd}}$.}
    \label{Fig: phase diagram-1}
\end{figure*}

For $m = 1$, $G_L^{(1)}(z)$ is
\begin{align}
    G_L^{(1)} = \frac{z-1+A}{z-1}.\label{GC1}
\end{align}
The above may be written in the continued fraction representation $G_C^{(1)}(z,\theta^{(1)})$ with the following Krylov angles
\begin{align}
    \cos\theta^{(1)}_k = \frac{(-1)^{k-1}A}{1+(k-1)A}.\label{Eq: m=1 Krylov angle}
\end{align}
The proof is presented in Appendix \ref{Sec: B} employing two methods. One is a brute force check on Mathematica. The second is using \eqref{Eq: m=1 Krylov angle} to derive the truncated continued fraction $G_C(z,\theta_k^{(1)};M)$ and showing that in the limit of $M\rightarrow \infty$, the expression agrees with \eqref{GC1}.

For $m = 2$, $G_L^{(2)}(z)$ is
\begin{align}
    G_L^{(2)} = \frac{-z-1+A}{-z-1}.
\end{align}
It is related to the $m = 1$ case by reflection, $G_L^{(2)}(z) = G_L^{(1)}(-z)$. The reflection in $z$ leaves the even Krylov angles unchanged, while transforming the odd Krylov angles to their supplementary angles (see Appendix \ref{Sec: C}). Namely, $G_C^{(2)}(z,\theta^{(2)})$ is parametrized by
\begin{align}
    \theta^{(2)}_k = \Big\{\begin{array}{cc}
    \pi - \theta^{(1)}_k & \text{if}\ k\ \text{is odd};\\
     \theta^{(1)}_k & \text{else}.
    \end{array}
    \label{Eq: m=2 Krylov angle}
\end{align}
For $m = 4$, $G_L^{(4)}$ is given by
\begin{align}
    G_L^{(4)}(z) = \frac{-z^2-1+A}{-z^2-1}.
\end{align}
It is invariant under reflection, $G_L^{(4)}(z) = G_L^{(4)}(-z)$, and related to $m = 1$ by square-reflection, $G_L^{(4)}(z) = G_L^{(1)}(-z^2)$. Due to the reflection invariance, all odd Krylov angles of $G_C^{(4)}(z,\theta^{(4)})$ are $\pi/2$ as $\pi/2$ is its own supplementary angle. The even angles of $\theta^{(4)}$ are mapped one-to-one from $\theta^{(1)}$ according to square-reflection, see Appendix \ref{Sec: D}. In summary the Krylov angles of  $G_C^{(4)}(z,\theta^{(4)})$ obey
\begin{align}
    \theta^{(4)}_k = \Big\{\begin{array}{cc}
    \pi/2 & \text{if}\ k\ \text{is odd};\\
     \theta^{(1)}_{k/2} & \text{else}.
    \end{array}.
    \label{Eq: m=4 Krylov angle}
\end{align}

The Floquet-ITFIM \eqref{ITFIM} is a classic model of topological physics 
in 1D \cite{SondhiRev}. Fig.~\ref{Fig: phase diagram-1} shows the topological phase diagram for  homogeneous couplings where all the odd angles are equal  $\theta_1 = \theta_3 = \theta_5 = \ldots = \theta_{\text{odd}}$ (uniform local transverse field) and all the even angles are equal $\theta_2 = \theta_4 = \theta_6 = \ldots = \theta_{\text{even}}$ (uniform Ising coupling). There are four phases that are characterized by the nature of the stable edge modes. These phases are a trivial-phase (no topologically protected edge modes), $0 (\pi)$-mode phase where a zero (period doubled $\pi$) mode resides at the edge and a $0-\pi$ mode phase where the edge hosts both a zero mode and a $\pi$ mode. The Floquet operator Krylov space returns spatially inhomogenous Krylov angles. Nevertheless, it is illuminating to study where these Krylov angles lie on the phase diagram.

\begin{figure*}
    \includegraphics[width=0.32\textwidth]{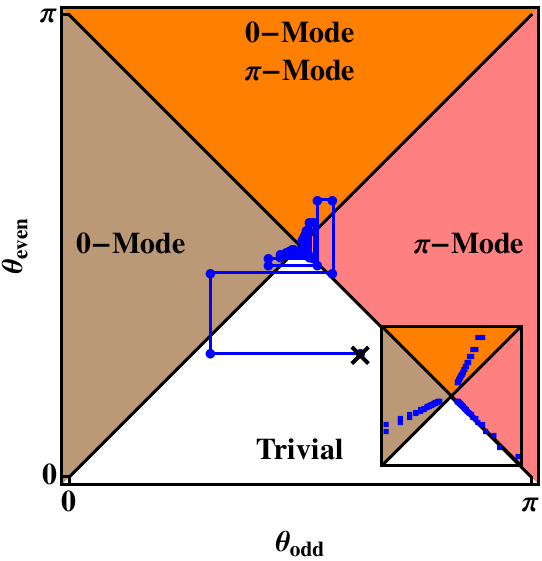}
    \includegraphics[width=0.32\textwidth]{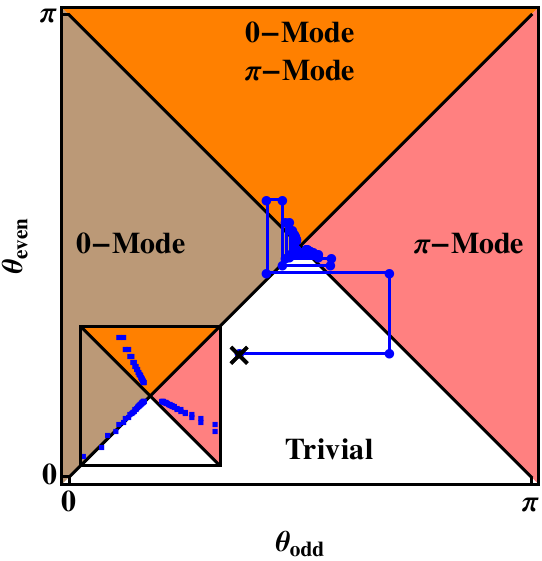}

    \caption{Same as Fig.~\ref{Fig: phase diagram-1} but showing the trajectories of Krylov angles for \eqref{Eq: m=3 Krylov angle}($m = 3$ period, left panel) and \eqref{Eq: m=6 Krylov angle} ($m=6$ period, right panel) with $A = 0.8$.  After period folding, these trajectories look similar to those of the left panel of Fig.~\ref{Fig: phase diagram-1}.}
    \label{Fig: phase diagram-2}
\end{figure*}

The Krylov angles solutions for $m =1,2, 4$ are presented in Fig.~\ref{Fig: phase diagram-1}. The trajectory of the Krylov angles is constructed by a series of data points: $\{ (\theta_1, \theta_2), (\theta_3, \theta_2), (\theta_3, \theta_4), (\theta_5, \theta_4), \ldots \}$. All trajectories approach the center of the phase diagram indicating that the bulk is critical. For $m = 1 (m = 2)$, the trajectory starts from the $0$-mode ($\pi$-mode)  phase (black cross in the left (middle) panel of Fig.~\ref{Fig: phase diagram-1}) and never leaves the $0$-mode ($\pi$-mode) phase. This is consistent with having stable 1-period (2-period) autocorrelation that oscillates with $0$ ($\pi$) phase every period. For $m = 4$, the trajectory starts from the trivial phase (black cross in the right panel of Fig.~\ref{Fig: phase diagram-1}) and approaches the critical point by traveling back and forth between the trivial and $0/\pi$-mode phase along the $\theta_{\text{odd}} = \pi/2$ vertical line. When $m > 2$, the phase diagram does not provide a good physical interpretation as the trajectories may travel across the trivial phase even for a non-decaying autocorrelation. One way to better interpret the results is by considering the period folded Krylov angles such that the original problem can be treated as an effective $0$-mode or $\pi$-mode problem \cite{yeh2023universal}. The 4-period autocorrelation can be folded into a 1-period or 2-period autocorrelation as follows: $A^{(4)}(4n) = A^{(1)}(n)$ or $A^{(4)}(2n) = A^{(2)}(n)$ for $n$ being a non-negative integer. On folding, the Krylov angles on the right panel of Fig.~\ref{Fig: phase diagram-1} will be folded into Krylov angles in the left and middle panels in the same figure. In Ref.~\cite{yeh2023universal}, the period folding method was applied to the numerical study of the $Z_3$ clock model \cite{sreejith2016parafermion,Fendley2012,jermyn2014stability}  where the period folded trajectory resembles the left panel in Fig.~\ref{Fig: phase diagram-1}. See Ref.~\cite{yeh2023universal} for results that support a topological interpretation of the autocorrelation function. In particular, the behavior of the 
autocorrelation can be interpreted as the edge dynamics of a topologically trivial or non-trivial chain attached to a chaotic bulk. The long time behavior is qualitatively different depending on where the local couplings fall in the topological phase diagram.

\subsection{$3,6$ period autocorrelation}

For $m = 3$, the discrete Laplace transform  $G_L^{(3)}(z)$ is 
\begin{align}
    G_L^{(3)}(z) = \frac{1-A + (1-\frac{A}{2})z+z^2}{1+z+z^2}.
\end{align}
The corresponding Krylov angles of the continued fraction representation $G_C^{(3)}(z,\theta^{(3)})$ are
\begin{subequations}\label{Eq: m=3 Krylov angle}
\begin{align}
    &\cos\theta^{(3)}_{3k-2} = \frac{(-1)^kA}{2+3(k-1)A},\\
    &\cos\theta^{(3)}_{3k-1} = \frac{(-1)^{k-1}A}{2+3(k-1)A - A},\\
    &\cos\theta^{(3)}_{3k} = \frac{(-1)^{k-1}2A}{2+3(k-1)A + A}, 
\end{align}
\end{subequations}
where $k$ is a positive integer. The proof is presented in Appendix \ref{Sec: E}.

For $m = 6$, $G_L^{(6)}(z)$ is 
\begin{align}
    G_L^{(6)}(z) = \frac{1-A - (1-\frac{A}{2})z+z^2}{1-z+z^2}.
\end{align}
Similar to the relation between the $m = 1,2$ cases, we have $G_L^{(6)}(z) = G_L^{(3)}(-z)$. Therefore, the Krylov angles of $G_C^{(6)}(z,\theta^{(6)})$ satisfy
\begin{align}
    \theta^{(6)}_k = \Big\{\begin{array}{cc}
    \pi - \theta^{(3)}_k & \text{if}\ k\ \text{is odd};\\
     \theta^{(3)}_k & \text{else}.
    \end{array}
    \label{Eq: m=6 Krylov angle}
\end{align}

Similar to Fig.~\ref{Fig: phase diagram-1}, we present Krylov angle solutions for $m =3, 6$ in Fig.~\ref{Fig: phase diagram-2}. These two solutions can be related to the 1 or 2-period autocorrelation via the period folding method. For example, with $n$ being an integer, $A^{(3)}(3n) = A^{(1)}(n)$ for 3-period case and $A^{(6)}(6n) = A^{(1)}(n)$, $A^{(6)}(3n) = A^{(2)}(n)$ for the 6-period case. Therefore, the Krylov angles in Fig.~\ref{Fig: phase diagram-2} can be folded into Krylov angles in the left or middle panels in Fig.~\ref{Fig: phase diagram-1}.  Last, it is only for the case where the autocorrelation function is of the simple form $A\cos(2\pi n/m)$, that the $m=1,2,4$ and $m=3,6$ autocorrelators are related by the above mentioned reflection and square reflection transformations. This does not generalize to general $m$-period correlators.

\subsection{Power-law localized edge mode in Krylov space}

\begin{figure*}
    \includegraphics[width=0.32\textwidth]{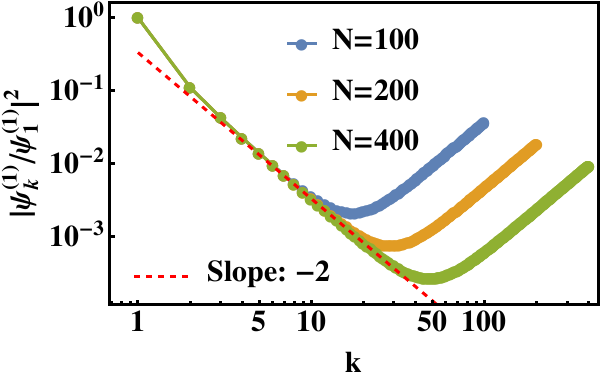}
    \includegraphics[width=0.32\textwidth]{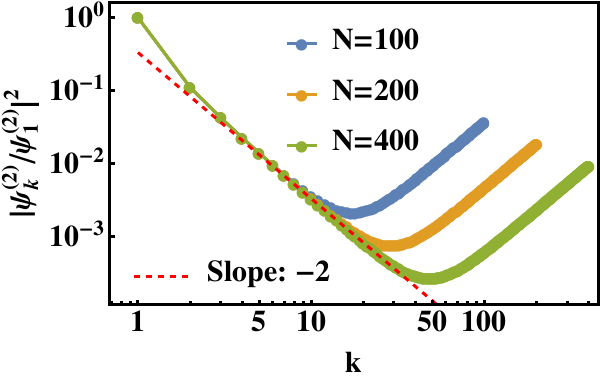}
    \includegraphics[width=0.32\textwidth]{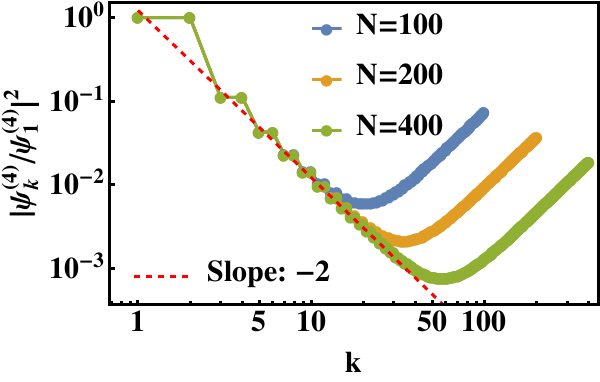}
    \includegraphics[width=0.32\textwidth]{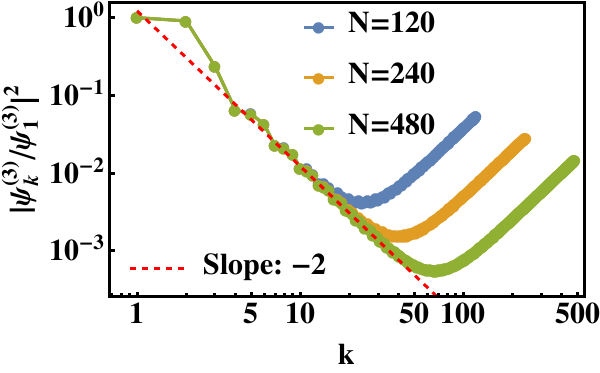}
    \includegraphics[width=0.32\textwidth]{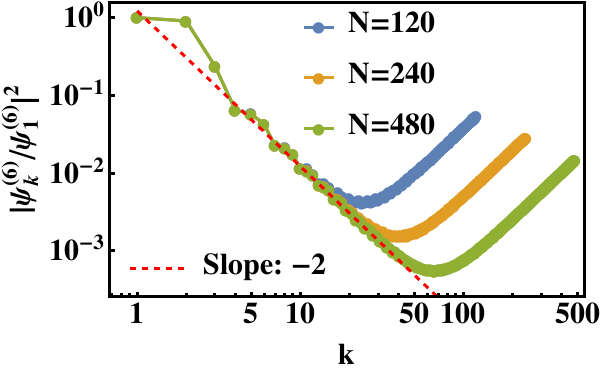}    

    \caption{The ED results for the edge modes from diagonalizing \eqref{Eq: K matrix Krylov basis K6}.  The $m$-period edge mode with $m = 1,2,4$ with system sizes $N = 100, 200, 400$ (top panels), and for $m=3,6$ with system sizes $N = 120, 240, 480$ (bottom panels) are shown. The choice of different system sizes for the top and bottom panels is to account for the difference in the underlying sub-lattice structure, see text. All the numerical results show the same power-law localization $\sim 1/k^2$ near the edge. $\psi^{(m)}$ is localized on both ends of the chain and is responsible for the system size effect corresponding to the upturn at large $k$.}
    \label{Fig: ED localization}
\end{figure*}

The persistent, i.e, non-decaying $m$-period autocorrelations is a consequence of the underlying $m$-period edge mode $\psi^{(m)}$ of the ITFIM constructed out of the Krylov angles. In the thermodynamic limit, $\psi^{(m)}$ satisfies the eigenoperator equation in the Krylov orthonormal basis: $\Tilde{K}\psi^{(m)} = e^{i2\pi/m}\psi^{(m)}$. Note that the Hermitian conjugation, $\psi^{(m)\dagger}$, is also an eigenoperator of the system with eigenvalue $e^{-i2\pi/m}$. Therefore, the existence of both edge modes $\psi^{(m)}$ and $\psi^{(m)\dagger}$ leads to the $\cos(2\pi n/m)$ oscillation of the autocorrelation. However, the Krylov angles \eqref{Eq: m=1 Krylov angle},\eqref{Eq: m=3 Krylov angle} lead to a Krylov chain with a critical bulk since $\theta_k \rightarrow \pi/2$ with $k \rightarrow \infty$.
This suggests non-exponentially localized edge modes in the system in contrast to the gapped system where for the latter the size of the gap sets the scale over which a mode is localized. For a gapless system, no such clear length scale exists. Consistent with this view, we now show that the edge modes are localized as a power-law. 

For $m=1$, using \eqref{Eq: m=1 Krylov angle},
one can show that the $1$-period edge mode (also known as the zero mode) in the Krylov space orthonormal basis \cite{yeh2023universal}, $\psi^{(1)} =(\psi_1^{(1)}, \psi_2^{(1)}, \ldots)^\intercal$ obeys the recursion relation
\begin{align}
    \left|\frac{\psi_{2l+1}^{(1)}}{\psi_{2l-1}^{(1)}}\right|^2&= \left(\tan\frac{\theta_{2l-1}^{(1)}}{2}\cot\frac{\theta_{2l}^{(1)}}{2}\right)^2\nonumber\\
    &= \frac{1+(2l-3)A}{1+(2l-1)A}\times\frac{1+(2l-2)A}{1+2lA}.
\end{align}
Using Mathematica, the asymptotic behavior of $\psi$ for large $k$ is
\begin{align}
    \left|\frac{\psi_{2k+1}^{(1)}}{\psi_{1}^{(1)}}\right|^2 &= \prod_{l=1}^k \frac{1+(2l-3)A}{1+(2l-1)A}\times\frac{1+(2l-2)A}{1+2lA}\nonumber\\
    &= \frac{1-A}{(1+2Ak)(1-A+2Ak)} \sim \frac{1-A}{4A^2}\cdot\frac{1}{k^2}.
\end{align}
Since Krylov space is one dimensional, $\psi^{(1)}$ is normalizable as long as $|\psi_{2k+1}^{(1)}|^2$ decays faster than $1/k$.  This implies that the zero mode in this example is power-law localized in Krylov space. The $m$-period autocorrelation can be viewed as a $1$-period autocorrelation after $m$-period folding \cite{yeh2023universal}, i.e, one only observes the autocorrelation at $n = ml$ for $A^{(m)}(n)$ where  $l$ is a non-negative integer. Therefore, the $m$-period edge mode is also power-law localized $\sim 1/k^2$ but with a different prefactor.

We present the numerical ED results for the $m$-period edge mode in Krylov space. This is obtained from diagonalizing \eqref{Eq: K matrix Krylov basis K6} for different system sizes $N$, and for the appropriate Krylov angles for the $m$-period autocorrelation. The eigenvector for the $m$-period edge mode $\psi^{(m)}$ is plotted in Fig.~\ref{Fig: ED localization}. Due to finite system size, the $\psi^{(m)}$ is the eigenvector with eigenvalue closest to $e^{i2\pi/m}$. For $m =1,2$, there are two eigenvalues closest to $e^{i2\pi/m}$ and the corresponding two eigenvectors are complex conjugates to each other. Therefore, choosing either one does not change the results in Fig.~\ref{Fig: ED localization}. The numerical results support that all the $m$-period edge modes obey the inverse-square power-law localization. 

We choose different system sizes for $m =1, 2, 4$ and $m = 3, 6$ to account for the underlying sub-lattice structure of the ITFIM. For the homogeneous transverse field Ising model, the system has two sub-lattices in the Majorana basis. The ITFIM for $m=1,2$ corresponding to the Krylov angles \eqref{Eq: m=1 Krylov angle} and \eqref{Eq: m=2 Krylov angle} respectively, can be thought of as a gradual deformation of the homogeneous model that hosts $0,\pi$ modes, such that the sub-lattice structure is the same. However, the $m=4$-period case \eqref{Eq: m=4 Krylov angle} corresponds to the insertion of a $\pi/2$ Krylov angle between every two $\theta^{(1)}$. Effectively, the sub-lattice is doubled. The $m=3,6$-period cases where the Krylov angles are \eqref{Eq: m=3 Krylov angle} and \eqref{Eq: m=6 Krylov angle} respectively, are equivalent to combining three ITFIMs alternatively, and the sub-lattice is tripled. This observation can also be seen in Figures \ref{Fig: phase diagram-1} and \ref{Fig: phase diagram-2}, where the trajectory does not approach the center from any unique direction for $m =3,4,6$ implying a bigger sub-lattice structure. Therefore, in Fig.~\ref{Fig: ED localization}, we choose system sizes divisible by the sub-lattice size.

\section{Numerical construction of the Krylov angles for decaying autocorrelations} \label{sec:Results2}

\begin{figure*}
    \includegraphics[width=0.32\textwidth]{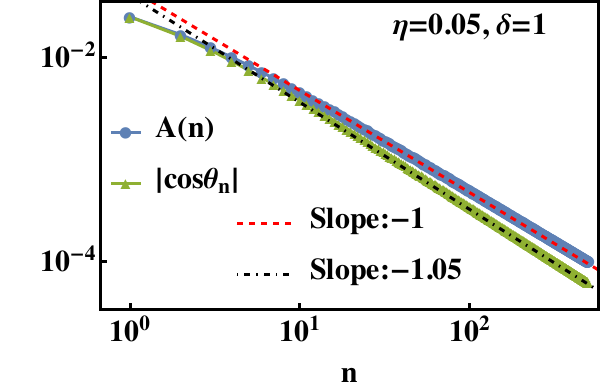}
    \includegraphics[width=0.32\textwidth]{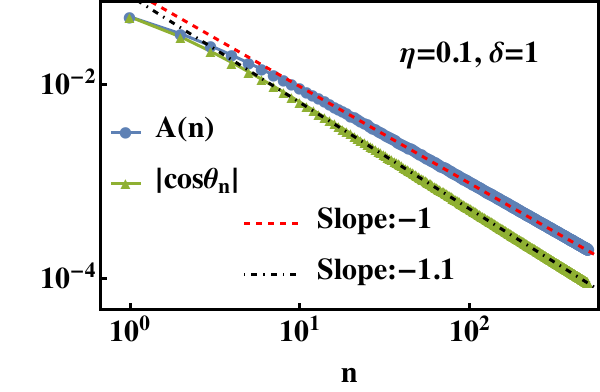}
    \includegraphics[width=0.32\textwidth]{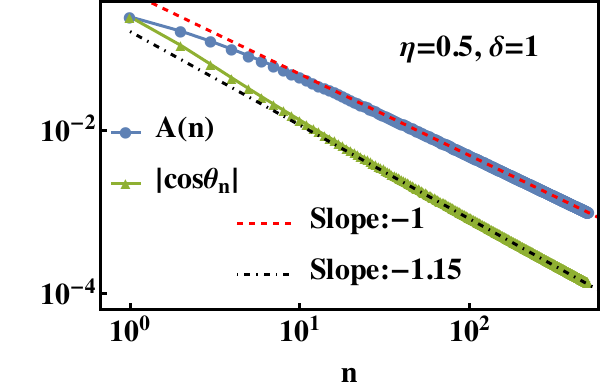}
    \includegraphics[width=0.32\textwidth]{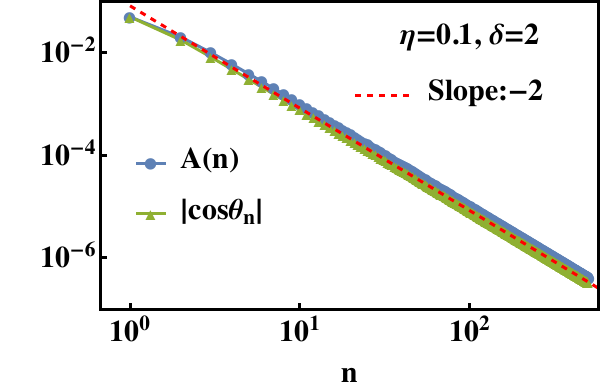}
    \includegraphics[width=0.32\textwidth]{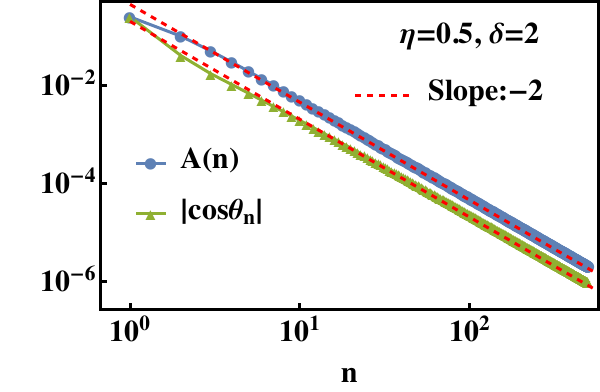}
    \includegraphics[width=0.32\textwidth]{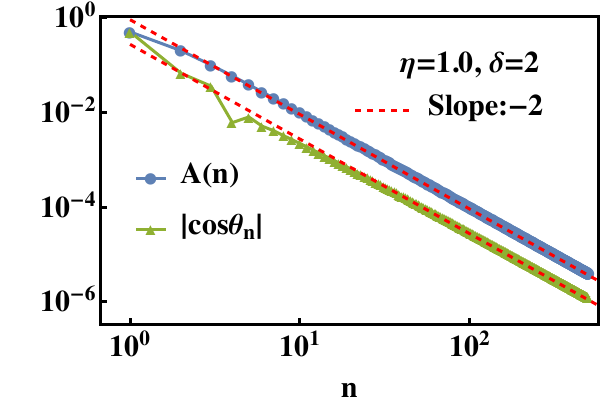}

    \caption{The Krylov angles are numerically computed from the power-law decaying autocorrelation \eqref{Eq: power-law decaying A} with the two plotted vs $n$. Here, $n$ represents stroboscopic time for $A(n)$ or the Krylov lattice index for $|\cos\theta_n|$. $|\cos\theta_n|$ shows the same power-law behavior as the autocorrelation for large $n$ for $\delta = 2$ (second row), with $\eta$ not altering the power-law. For $\delta = 1$ (first row), there is a difference between the power-law decays of the autocorrelation and the Krylov angles at large $n$, with the difference increasing with increasing $\eta$.}
    \label{Fig: Krylov power-law}
\end{figure*}

\begin{figure*}
    \includegraphics[width=0.32\textwidth]{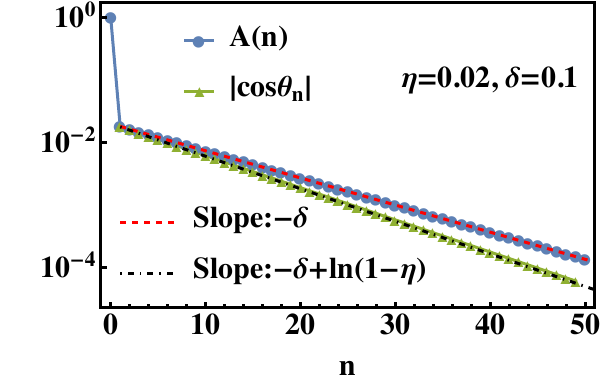}
    \includegraphics[width=0.32\textwidth]{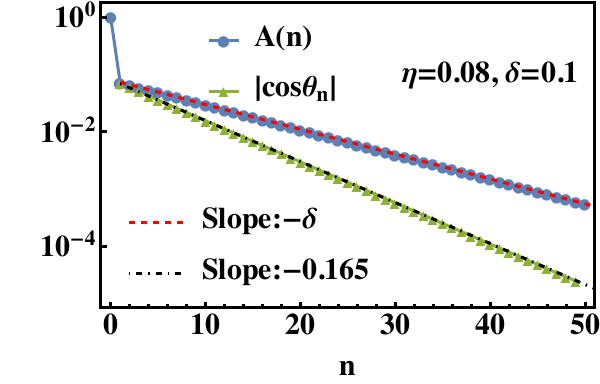}
    \includegraphics[width=0.32\textwidth]{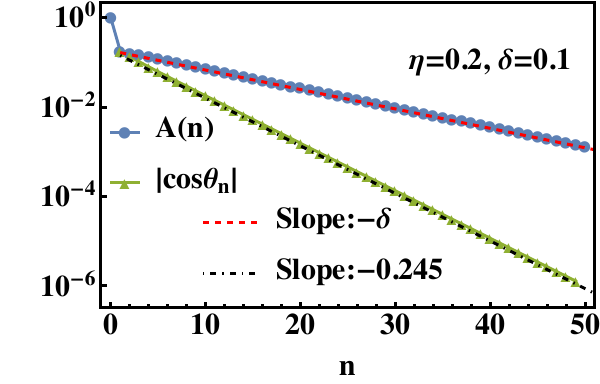}
    \includegraphics[width=0.32\textwidth]{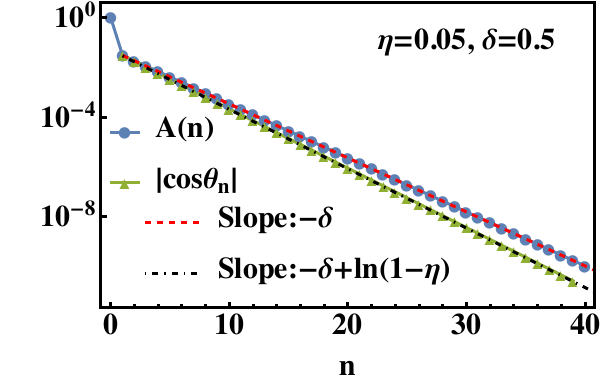}
    \includegraphics[width=0.32\textwidth]{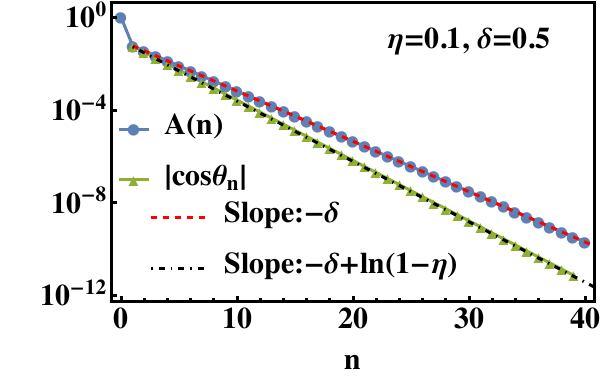}
    \includegraphics[width=0.32\textwidth]{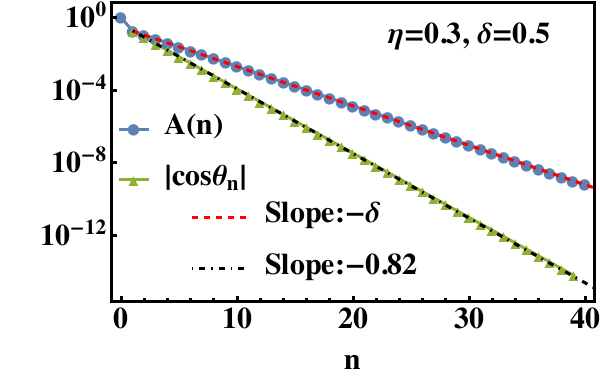}

    \caption{The Krylov angles are numerically computed from the exponentially decaying autocorrelation \eqref{Eq: exponential decaying A} with $\delta = 0.1$ (first row) and $\delta=0.5$ (second row). Both the autocorrelation and  $|\cos\theta_n|$ obey an exponential decay  at large $n$, with $|\cos\theta_n| \sim e^{-[\delta -\ln(1-\eta)]n}$  when $\eta \ll \delta$ (left panels). When $\eta$ becomes comparable to or larger than $\delta$, $|\cos\theta_n|$ still decays faster than the autocorrelation but the exponent of the exponential decay is not exactly $\left[\delta -\ln(1-\eta)\right]$ (middle and right panels).}
    \label{Fig: Krylov exponential}
\end{figure*}

In the previous section, we have presented analytic solutions for the Krylov angles for persistent, i.e, non-decaying in time, $m$-period autocorrelations with $m = 1,2,3,4,6$. In this section, we consider two cases of autocorrelations that decay in time: (i) power-law decaying autocorrelations,
\begin{align}
   A(n) = \frac{\eta}{1+n^\delta},
   \label{Eq: power-law decaying A}
\end{align}
and (ii) exponentially decaying autocorrelations, 
\begin{align}
   A(n) = \eta e^{-\delta n}.
   \label{Eq: exponential decaying A}
\end{align}
Above, $\eta$ and $\delta$ are free parameters. According to the algorithm proposed in Section \ref{Sec: algorithm}, the corresponding Krylov angles are numerically determined with some choice of $\eta$ and $\delta$.

For the power-law decaying correlator, i.e, case (i), the numerical results for $\delta = 1, 2$ are presented in Fig.~\ref{Fig: Krylov power-law}. The numerically computed $\cos\theta_n$ oscillate as $(-1)^{n-1}$, similar to \eqref{Eq: m=1 Krylov angle}. Thus  $|\cos\theta_n|$ is plotted in Fig.~\ref{Fig: Krylov power-law} along with the autocorrelation, where the index $n$ represents strobscopic time for the latter, and the Krylov lattice index for the former. Since the correlators eventually decay to zero, $\theta_n$ approach $\pi/2$ at large $n$. Interestingly, for $\delta = 2$, $|\cos\theta_n|$ approaches $\pi/2$ as a power-law, and with the same power as the autocorrelation approaches zero. The value of $\eta$ does not change the asymptotic behavior.  In contrast, $\eta$ does affect the asymptotic behavior for $\delta = 1$. In fact, for $\delta > 1$, the asymptotic power-law is the same for both the autocorrelation and $|\cos\theta_n|$.
This numerical observation can be proven by studying the system close to the maximally ergodic point where the Krylov angles are close to $\pi/2$.

\begin{figure}[h!]
    \includegraphics[width=0.4\textwidth]{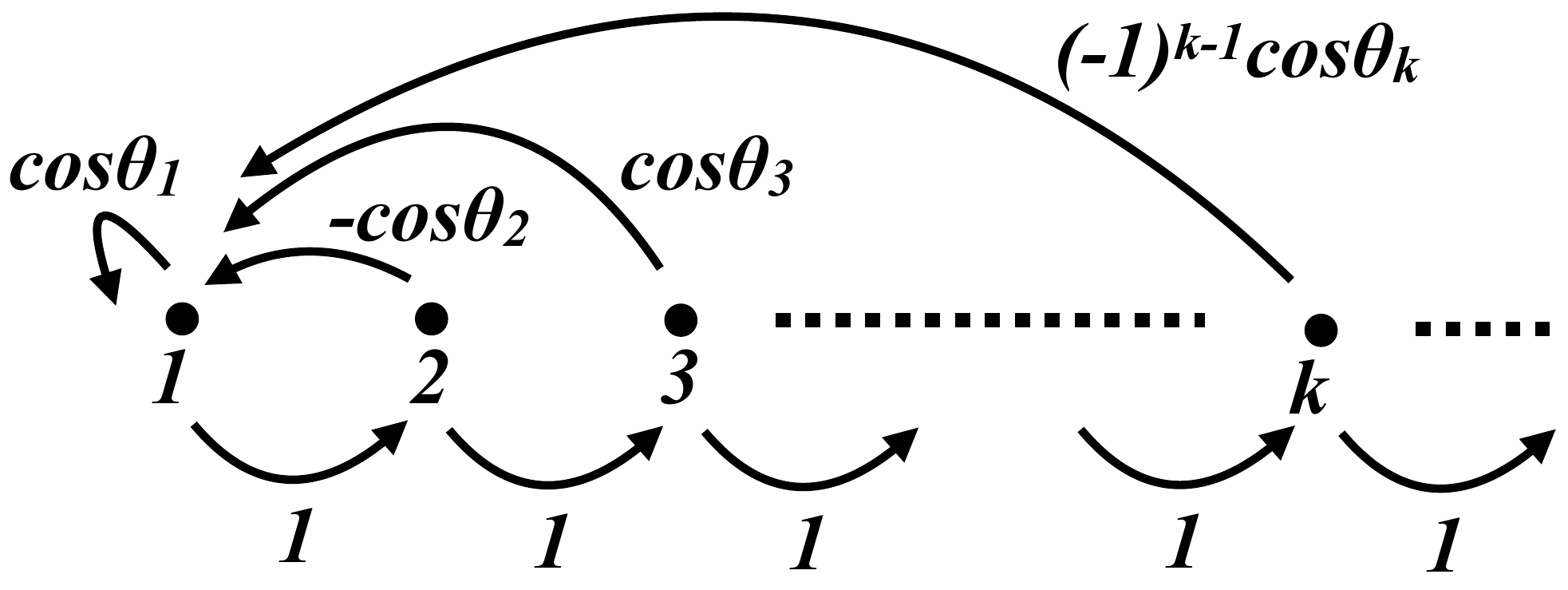}

    \caption{The pictorial illustration of \eqref{Eq: critical K tilde}. Around the maximally ergodic limit, the hopping in Krylov space is approximately either forward to the next site or backward all the way to the first site.}
    \label{Fig: critical hopping}
\end{figure}

Around the maximally ergodic limit, $\cos\theta_n$ is a small number and $\sin\theta_n$ is approximated by 1. In this limit, it is easier to work in the Krylov orthonormal basis \eqref{Eq: K-matrix Krylov space} where $\Tilde{K}$-matrix can be approximated as follows at $O(\cos\theta_n)$ 
\begin{align}
    \Tilde{K} \approx \begin{pmatrix}
        \cos\theta_1 & -\cos\theta_2 & \cos\theta_3 & -\cos\theta_4 & \cos\theta_5 & \ldots\\
        1 & 0 & 0 & 0 & 0 & \ldots\\
        0 & 1 & 0 & 0 & 0 & \ldots\\
        0 & 0 & 1 & 0 & 0 & \ddots\\
        0 & 0 & 0 & 1 & 0 & \ddots\\
        \vdots & \vdots & \vdots & \ddots & \ddots & \ddots
    \end{pmatrix}.
    \label{Eq: critical K tilde}
\end{align}
The approximate autocorrelation $\Tilde{A}(n) = (1|\Tilde{K}^n|1)$ has the following simple form
\begin{widetext}
\begin{align}
    \Tilde{A}(n) =& (-1)^{n-1}\cos\theta_n + \sum_{\substack{j_1,j_2 = 1\\ j_1 + j_2 = n}}^{n} (-1)^{n-2}\cos\theta_{j_1} \cos\theta_{j_2} + \sum_{\substack{j_1,j_2,j_3 = 1\\ j_1+j_2+j_3 = n}}^{n} (-1)^{n-3}\cos\theta_{j_1} \cos\theta_{j_2}\cos\theta_{j_3} + \ldots \nonumber\\
    &+ \sum_{\substack{j_1,j_2, \ldots, j_m = 1\\ j_1+j_2+\ldots+j_m = n}}^{n} (-1)^{n-m}\cos\theta_{j_1} \cos\theta_{j_2}\ldots\cos\theta_{j_m} + \ldots . \label{eq:exp1}
\end{align}
\end{widetext}
The above expansion counts how many times the particle returns to the first site within $n$ time steps. The hopping rules are summarized in Fig.~\ref{Fig: critical hopping}, where the particle is only allowed to go to the next site to the right, or return to the first site. The first term in the expansion of  \eqref{eq:exp1} corresponds to the particle traveling to the $n$-th site and then hopping back to the first site, which takes $n$ time steps. The second term in \eqref{eq:exp1} corresponds to the particle going to the $j_1$-th site and back to the first site with amplitude $(-1)^{j_1-1}\cos\theta_{j_1}$. Then, it travels to the $j_2$-th site and returns to the first site with amplitude $(-1)^{j_2-1}\cos\theta_{j_2}$, where in total it takes $j_1 + j_2$ time steps. Since we require that the total time steps is $n$, one has to sum over all possible $j_1$ and $j_2$ with the constraint $j_1 + j_2 = n$. In general, for the $m$-th term, the particle comes back to the first site $m$ times from sites $j_1, j_2, \ldots, j_m$ separately with the constraint that $j_1+j_2+ \ldots +j_m = n$, which corresponds to the second line of \eqref{eq:exp1}. 

We assume $\cos\theta_n$ obeys the power-law behavior: $\cos\theta_n = (-1)^{n-1}\eta/n^{\delta}$, where $\eta \ll 1$ such that $|\cos\theta_n| \ll 1$. The approximate autocorrelation is
\begin{align}
    \Tilde{A}(n) &\sim \frac{\eta}{n^\delta} + \eta^2\sum_{\substack{j_1,j_2 = 1\\ j_1 + j_2 = n}}^{n}\frac{1}{{j_1}^\delta {j_2}^\delta} + \ldots \nonumber\\
    &= \frac{\eta}{n^\delta} + \eta^2\sum_{j_1 = 1}^{n-1}\frac{1}{j_1^\delta(n-j_1)^\delta} + \ldots
\end{align}
We can estimate the asymptotics for large $n$ by replacing the summation in the second term by an integral.
\begin{align}
    \sum_{j_1 = 1}^{n-1}\frac{1}{j_1^\delta(n-j_1)^\delta} \sim \int_{1}^{n-1}\frac{dx}{x^\delta(n-x)^\delta} \sim \Big\{\begin{array}{cc}
     \frac{1}{n^\delta} & \text{if } \delta > 1;\\
     \frac{\ln{n}}{n^\delta} &\text{if } \delta = 1. \end{array}
\end{align}
Therefore, $A(n) \sim 1/n^\delta$ has the same asymptotic form as $|\cos\theta_n|$ for $\delta >1$ since the higher order summation terms will not change the power-law behavior but only modify the prefactor. However, this relation fails for $\delta = 1$ due to the logarithmic correction. Qualitatively, the numerics indicate that to obtain the $1/n$ asymptotics of the autocorrelation, one needs $\cos\theta_n$ to decay slightly faster than $1/n$, but not as a simple power-law, to cancel the logarithmic correction.

\begin{figure}[h!]
    \includegraphics[width=0.4\textwidth]{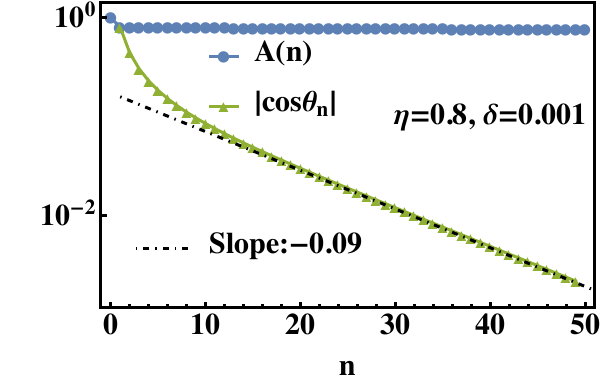}

    \caption{The Krylov angles are computed from exponentially decaying autocorrelation \eqref{Eq: exponential decaying A} with $\eta = 0.8$ and $\delta=0.001$. The corresponding $|\cos\theta_n|$ decays faster than $A(n)$ so that it can be a sensitive probe of a small decay rate $\delta \ll 1$. }
    \label{Fig: slow exponential decaying}
\end{figure}

Now we discuss the Krylov angles for an exponentially decaying correlator, case (ii) above. The numerical results are shown in Fig.~\ref{Fig: Krylov exponential} for $\delta = 0.1, 0.5$. The numerical observation is that when $\eta \ll \delta$, the asymptotics of $|\cos\theta_n|$ follow $e^{-[\delta -\ln(1-\eta)]n}$. As $\eta$ becomes comparable or larger than $\delta$, the exponent is close to but not precisely $\left[\delta -\ln(1-\eta)\right]$. This result can also be obtained analytically by studying the Krylov angles around the maximally ergodic limit of $\theta=\pi/2$. Substituting $\cos\theta_n = (-1)^{n-1}\mu e^{-{\nu n}}$ in \eqref{eq:exp1}, the approximate autocorrelation is
\begin{align}
    \Tilde{A}(n) =& \mu e^{-\nu n} + \mu^2\sum_{\substack{j_1,j_2 = 1\\ j_1 + j_2 = n}}^{n}  e^{-\nu n} \nonumber\\
    &+ \mu^3\sum_{\substack{j_1,j_2,j_3 = 1\\ j_1+j_2+j_3 = n}}^{n} e^{-\nu n} + \ldots \nonumber\\
    &+ \mu^{m}\sum_{\substack{j_1,j_2, \ldots, j_m = 1\\ j_1+j_2+\ldots+j_m = n}}^{n} e^{-\nu n} + \ldots.
\end{align}
The summations lead to binomial coefficients \cite{sedgewick2009analytic}, 
\begin{align}
    \sum_{\substack{j_1,j_2, \ldots, j_m = 1\\ j_1+j_2+\ldots+j_m = n}}^{n} 1 = \binom{n-1}{m-1}.
\end{align}
The the approximate autocorrelation is therefore 
\begin{align}
    \Tilde{A}(n) &= \mu e^{-\nu n} \left[ \sum_{m=1}^{n} \binom{n-1}{m-1} \mu^{m-1} \right]\nonumber\\
    &=\mu(\mu+1)^{n-1} e^{-\nu n}\nonumber\\
    &=\frac{\mu}{\mu+1}e^{-[\nu - \ln(\mu+1)]n}.
\end{align}
By comparing with \eqref{Eq: exponential decaying A}, one concludes that $\mu = \eta/(1-\eta)$ and $\nu = \delta - \ln(1-\eta)$.
Thus $A(n) \propto e^{-\delta n}$ and $|\cos\theta_n| \propto e^{-[\delta -\ln(1-\eta)]n}$ for large $n$.
A numerical observation is that this results holds for $\eta \ll \delta$. Note that $|\cos\theta_n|$ still decays faster than $A(n)$ for $\eta > \delta$ as shown in Fig.~\ref{Fig: Krylov exponential}. Even if one considers a slowly decaying autocorrelation, $\delta \ll 1$, $|\cos\theta_n|$ will still decay exponentially with a larger exponent. Thus Krylov angles are sensitive to long-lived autocorrelations with a small decay rate, see Fig.~\ref{Fig: slow exponential decaying} for an illustration.  

Although the approximate result $|\cos\theta_n| \sim e^{-[\delta -\ln(1-\eta)]n}$ is valid for $\eta \ll \delta$, it correctly indicates that $\eta = 1$ is a special case with $\cos\theta_{n\geq 2} = 0$. To show this, we first explicitly compute $\cos\theta_1$ from \eqref{Eq: A(1)}.
\begin{align}
    &\cos\theta_1 = A(1) = e^{-\delta}.
\end{align}
Note that $A(n) = e^{-\delta n}$ and $A(n) = (1|\Tilde{K}^n|1)$. Therefore, one concludes that $A(n) = \cos^n\theta_1 = (\Tilde{K}_{11})^n$, where $\Tilde{K}_{11}$ is the upper most left matrix element of $\Tilde{K}$-matrix. In summary, the only trajectories are those where the particle either stays on the first site or hops away from it never returning to the first site. This implies $\cos\theta_{n\geq 2} = 0$, equivalently $\theta_{n\geq 2} =\pi/2$. This is equivalent to a dot attached to a maximally ergodic Krylov chain. In fact, this special case was discussed in Ref.~\cite{suchsland2023krylov}, in order to highlight that the Krylov space is sensitive to the underlying maximal ergodicity regardless of the value of $\delta$.

\section{Conclusions} \label{Sec: Conc}
We proved a moment method where the Krylov angles parametrizing a Floquet operator Krylov space can be obtained directly from the Floquet infinite temperature autocorrelation function. This algorithm allows one to bypass the computationally expensive Gram-Schmidt procedure for constructing the operator Krylov space. The algorithm also shows that there is a direct relation between the number of Krylov angles and the stroboscopic time upto which the dynamics is known. Moreover, the algorithm sets constraints on autocorrelations that can be generated by unitary dynamics. 

We also derived the continued fraction representation of the discrete Laplace transformation. The Krylov angles naturally appear in the parameters of the continued fraction and this connection leads to an analytic approach for solving for Krylov angles. We presented  analytic solutions for the Krylov angles for persistent $m$-period autocorrelations for some simple forms of the autocorrelation functions. The dynamics corresponded to power-law localized $m$-period edge modes in gapless 1D Floquet systems. We also presented numerical studies on power-law and exponentially decaying autocorrelations. The relation between decaying dynamics and Krylov angles was obtained by a perturbative calculation near the critical point $\theta_i=\pi/2$, the latter being equivalent to the maximally ergodic dual-unitary limit \cite{suchsland2023krylov}.

There are many directions for future research. The Floquet operator Krylov mapping to a 1D ITFIM indicates a practical avenue for engineering Floquet systems with desired edge modes because the mapping is between a seed operator and an edge mode of the ITFIM. It would also be interesting to explore how the operator growth hypothesis for Hamiltonian dynamics \cite{parker2019universal}   generalizes to Floquet dynamics.

{\sl Acknowledgments:} 
This work was supported by the US Department of Energy, Office of
Science, Basic Energy Sciences, under Award No.~DE-SC0010821. HCY acknowledges support of the NYU IT High Performance Computing resources, services, and staff expertise.

\appendix
\section{Derivation of continued fraction formula \eqref{Eq: G continued fraction}, \eqref{eq:alpha-beta}}
\label{Sec: A}
For the ITFIM, we consider the time evolution of  the operator expanded in the Majorana basis $\{ \gamma_1,\ldots,\gamma_N\}$
\begin{align}
    \psi(n) = \sum_{k=1}^{N} \psi_k(n) \gamma_k,
\end{align}
where the initial conditions are: $\psi_1(0) = 1, \psi_{k\neq 1}(0) = 0$. In the stroboscopic time step from $n$ to $n+1$ $\psi_k(n)$ evolve as follows: 
\begin{subequations}
\begin{align}
    \psi_1 (n+1) =& \cos\theta_1  \psi_1(n) + \sin\theta_1  \psi_2(n);\label{Eq: one step evolution-1}\\
    \psi_{2l} (n+1) =& -\sin\theta_{2l-1}\cos\theta_{2l}  \psi_{2l-1}(n) \nonumber\\
    &+ \cos\theta_{2l-1}\cos\theta_{2l} \psi_{2l}(n) \nonumber\\  
    &+ \sin\theta_{2l}\cos\theta_{2l+1}  \psi_{2l+1}(n) \nonumber\\
    &+ \sin\theta_{2l}\sin\theta_{2l+1} \psi_{2l+2} (n);\label{Eq: one step evolution-2}\\
    \psi_{2l+1} (n+1) =& \sin\theta_{2l-1}\sin\theta_{2l} \psi_{2l-1}(n) \nonumber\\
    &- \cos\theta_{2l-1}\sin\theta_{2l} \psi_{2l}(n) \nonumber\\  
    &+ \cos\theta_{2l}\cos\theta_{2l+1} \psi_{2l+1}(n) \nonumber\\
    &+ \cos\theta_{2l}\sin\theta_{2l+1} \psi_{2l+2} (n);\label{Eq: one step evolution-3}\\
    \psi_N(n+1) =& -\sin\theta_{N-1} \psi_{N-1}(n) + \cos\theta_{N-1} \psi_N(n).\label{Eq: one step evolution-4}
\end{align}
\end{subequations}
Above we focus on $\psi_k(n+1)$ at $k = 1, 2l, 2l+1, N$ due to the repetitive structure of \eqref{Eq: K matrix}.
The discrete Laplace transformation of the autocorrelation function
$A_k(n)= {\rm Tr}\left[\gamma_k(n)\gamma_k(0)\right]/N$ corresponds to the Laplace transform of $\psi_k(n)$ which we denote by $G_k$ 
\begin{align}
    G_k(z) = \sum_{n=0}^{\infty} \psi_k(n) z^{-n}.
\end{align}
With the above definition, one can show that
\begin{align}
    \sum_{n=0}^{\infty} \psi_k(n+1) z^{-n} = z\left(-\psi_k(0) + G_k(z)\right).
\end{align}
Therefore the Laplace transformations of \eqref{Eq: one step evolution-1}, \eqref{Eq: one step evolution-2}, \eqref{Eq: one step evolution-3} and \eqref{Eq: one step evolution-4} become
\begin{subequations}
\begin{align}
    -z + zG_1 =& \cos\theta_1  G_1 + \sin\theta_1  G_2,\label{Eq: Floquet recursion-1}\\
    zG_{2l} =& -\sin\theta_{2l-1}\cos\theta_{2l}  G_{2l-1} \nonumber\\
    &+ \cos\theta_{2l-1}\cos\theta_{2l} G_{2l} \nonumber\\  
    &+ \sin\theta_{2l}\cos\theta_{2l+1}  G_{2l+1} \nonumber\\
    &+ \sin\theta_{2l}\sin\theta_{2l+1}  G_{2l+2},\label{Eq: Floquet recursion-2}\\
    zG_{2l+1} =& \sin\theta_{2l-1}\sin\theta_{2l}   G_{2l-1} \nonumber\\
    &- \cos\theta_{2l-1}\sin\theta_{2l}  G_{2l} \nonumber\\  
    &+ \cos\theta_{2l}\cos\theta_{2l+1}   G_{2l+1} \nonumber\\
    &+ \cos\theta_{2l}\sin\theta_{2l+1}  G_{2l+2},\label{Eq: Floquet recursion-3}\\
    zG_N =& -\sin\theta_{N-1} G_{N-1} + \cos\theta_{N-1} G_N,
\end{align} 
\end{subequations}
where we omit the $z$-dependence for simplicity and apply the initial conditions: $\psi_1(0) = 1, \psi_{k\neq 1}(0) = 0$.
To derive the recursion relation, we have to combine \eqref{Eq: Floquet recursion-2} and \eqref{Eq: Floquet recursion-3}. First, we consider the summation of \eqref{Eq: Floquet recursion-2} multiplied by $(-\cos\theta_{2l})$ and \eqref{Eq: Floquet recursion-3} multiplied by $\sin\theta_{2l}$,
\begin{align}
    &-\cos\theta_{2l} zG_{2l} + \sin\theta_{2l} z G_{2l+1} \nonumber\\
    &= \sin\theta_{2l-1} G_{2l-1} -\cos\theta_{2l-1} G_{2l}.
\end{align}
Rearranging the equation, one obtains
\begin{align}
    \frac{G_{2l}\sin\theta_{2l-1}}{G_{2l-1}} = \frac{-\sin^2\theta_{2l-1}}{z\cos\theta_{2l} - \cos\theta_{2l-1} -   \frac{G_{2l+1}}{G_{2l}}z\sin\theta_{2l}}.
\end{align}
Next, we consider the summation of \eqref{Eq: Floquet recursion-2} multiplied by $\sin\theta_{2l}$ and \eqref{Eq: Floquet recursion-3} multiplied by $\cos\theta_{2l}$,
\begin{align}
    &\sin\theta_{2l}  zG_{2l} + \cos\theta_{2l} zG_{2l+1} \nonumber\\
    &= \cos\theta_{2l+1}  G_{2l+1} + \sin\theta_{2l+1}  G_{2l+2}.\label{Eq: G recursion-1}
\end{align}
Rearranging the equation, one obtains
\begin{align}
    \frac{G_{2l+1}z\sin\theta_{2l}}{G_{2l}} = \frac{-z^2\sin^2\theta_{2l}}{ z\cos\theta_{2l} - \cos\theta_{2l+1} - \frac{G_{2l+2}\sin\theta_{2l+1}}{G_{2l+1}}}.\label{Eq: G recursion-2}
\end{align}
Now, we are ready to express $G_1$ as a continued fraction. By rearranging \eqref{Eq: Floquet recursion-1}, one obtains
\begin{align}
    G_1 = \frac{z}{z - \cos\theta_1 - \frac{G_2\sin\theta_1}{G_1}}. 
\end{align}
With \eqref{Eq: G recursion-1} and \eqref{Eq: G recursion-2}, one arrives at the continued fraction expression for $G_1$
\begin{align}
    G_1 = \frac{z}{z - \cos\theta_1 + \frac{\sin^2\theta_{1}}{z\cos\theta_2 - \cos\theta_1 + \frac{z^2\sin^2\theta_2}{z\cos\theta_2 -\cos\theta_3 +\frac{\sin^2\theta_3}{\ddots}}}}.
\end{align}
Since $\psi_1(n)$ corresponds to the autocorrelation, $A(n) = (1|K^n|1) = \psi_1(n)$, $G_1$ is the same as the $G_C(z,\theta)$ defined in \eqref{Eq: G continued fraction}.
We now prove some special properties of the truncated series
\begin{subequations}
\begin{align}
    &G_C(z,\theta; M) = \frac{\alpha_0}{\beta_0 + \frac{\alpha_1}{  \ddots+\frac{\alpha_M}{\beta_M}}} = \frac{P_M}{Q_M},\\  &\underset{M\rightarrow \infty}{\lim}G_C(z,\theta; M)= G_C(z,\theta),
    \label{Eq: truncated G}
\end{align}
\end{subequations}
where $P_M$ and $Q_M$ are polynomials of $z$ and therefore $P_M/Q_M$ is a rational function. We will show that they satisfy the recursive relations \cite{jones2009continued} 
\begin{align}
    P_k = \beta_k P_{k-1} + \alpha_k P_{k-2};\ Q_k = \beta_k Q_{k-1} + \alpha_k Q_{k-2},
\end{align}
where $k$ is a non-negative integer. Requiring that $P_0/M_0 = \alpha_0/\beta_0$ gives the initial conditions $P_{-2} = 1, P_{-1} = 0, Q_{-2} = 0$ and $Q_{-1} = 1$. The above recursive relations for $P$ and $Q$ can be proved by mathematical induction. We first assume that the recursive relation is true up to some number $M^*$, namely
\begin{align}
    G_C(z,\theta; M^*) &= \frac{\alpha_0}{\beta_0 + \frac{\alpha_1}{  \ddots+\frac{\alpha_{M^*}}{\beta_{M^*}}}}\nonumber\\
    &= \frac{P_{M^*}}{Q_{M^*}} = \frac{\beta_{M^*}P_{M^*-1}+\alpha_{M^*}P_{M^*-2}}{\beta_{M^*}Q_{M^*-1}+\alpha_{M^*}Q_{M^*-2}},
    \label{Eq: PQ M^*}
\end{align}
Next, we prove the recursive relation also holds for $M^* + 1$. For $M^*+1$, we have
\begin{align}
    G_C(z,\theta; M^*+1) = \frac{\alpha_0}{\beta_0 + \frac{\alpha_1}{  \ddots+\frac{\alpha_{M^*}}{\beta_{M^*}+ \frac{\alpha_{M^*+1}}{\beta_{M^*+1}}}}} = \frac{P_{M^*+1}}{Q_{M^*+1}}.
\end{align}
Note that one can treat $G_C(z,\theta;M^*+1)$ as $G_C(z,\theta;M^*)$ by replacing $\beta_{M^*}$ by $\beta_{M^*} + \alpha_{M^*+1}/\beta_{M^*+1}$. Therefore, by the last equality in \eqref{Eq: PQ M^*},
\begin{align}
    \frac{P_{M^*+1}}{Q_{M^*+1}} &= \frac{\left(\beta_{M^*}+ \frac{\alpha_{M^*+1}}{\beta_{M^*+1}}\right)P_{M^*-1}+\alpha_{M^*}P_{M^*-2}}{\left(\beta_{M^*}+ \frac{\alpha_{M^*+1}}{\beta_{M^*+1}}\right)Q_{M^*-1}+\alpha_{M^*}Q_{M^*-2}} \nonumber\\
    &= \frac{P_{M^*}+\frac{\alpha_{M^*+1}}{\beta_{M^*+1}}P_{M^*-1}}{Q_{M^*}+\frac{\alpha_{M^*+1}}{\beta_{M^*+1}}Q_{M^*-1}} \nonumber\\
    &= \frac{\beta_{M^*+1}P_{M^*}+\alpha_{M^*+1}P_{M^*-1}}{\beta_{M^*+1}Q_{M^*}+\alpha_{M^*+1}Q_{M^*-1}},
\end{align}
where we have used \eqref{Eq: PQ M^*} in the second equality. Now, we have proved that the recursive relation holds for $M^*+1$ if it also holds for $M^*$. By directly checking that the recursive relations of $P$ and $Q$ hold for $M^* = 0$ with boundary condition $P_{-2} = 1, P_{-1} = 0, Q_{-2} = 0$ and $Q_{-1} = 1$, the recursive relations automatically hold for general non-negative integer $M^*$ due to mathematical induction.

\section{Krylov angles \eqref{Eq: m=1 Krylov angle} of 1-period autocorrelation}
\label{Sec: B}

The analytic expression of the Krylov angles \eqref{Eq: m=1 Krylov angle} are conjectured from solving the first few angles analytically by equating $A^{(1)}(n) = (1|K^n|1)$. $A^{(1)}(n)$
obeys \eqref{Adef}, thus, $A^{(1)}(0)=1, A^{(1)}(n>0)=A$. Here we explicitly show the first few terms. For $n = 1$,
\begin{align}
    A^{(1)}(1) = \cos\theta_1^{(1)}.
\end{align}
Therefore, one obtains $\cos\theta_1^{(1)} = A$ and $\sin\theta_1^{(1)} = \sqrt{1-A^2}$. Note that the Krylov angles are defined between $[0,\pi]$ so that $\sin\theta_k^{(1)} \geq 0$ for all positive integers $k$. For $n = 2$,
\begin{align}
    A^{(1)}(2) &= \cos^2\theta_1^{(1)} -\sin^2\theta_1^{(1)} \cos\theta_2^{(1)}\nonumber\\
    &= A^2 - (1-A^2)\cos\theta_2^{(1)}.
\end{align}
Since $A^{(1)}(2)=A$, one obtains $\cos\theta_2^{(1)} = -A/(1+A)$ and $\sin\theta_2^{(1)} = \sqrt{1+2A}/(1+A)$. The higher terms can be obtained easily by symbolic computation in Mathematica, and we conjecture the analytic expression \eqref{Eq: m=1 Krylov angle} by observing the patterns of the first few terms. Direct simulations for the autocorrelation for $\gamma_1$ for the angles \eqref{Eq: m=1 Krylov angle}, supports the result.

The discrete Laplace transformation of $A^{(1)}(n)$ is
\begin{align}
    G_L^{(1)}(z)  = \frac{z-1+A}{z-1}.
\end{align}
An alternate way to justify the Krylov angles in \eqref{Eq: m=1 Krylov angle} is to
prove that the equivalent continued fraction representation $G_C^{(1)}(z,\theta^{(1)})$ is parametrized by the Krylov angles \eqref{Eq: m=1 Krylov angle}. For this, we consider the truncated continued fraction $G_C^{(1)}(z,\theta^{(1)}; M)$,  defined as
\begin{align}
    G_C^{(1)}(z,\theta^{(1)}; M) = \frac{\alpha_0}{\beta_0 + \frac{\alpha_1}{  \ddots+\frac{\alpha_M}{\beta_M}}} = \frac{P_M}{Q_M}.
\end{align}
Now our goal is to show that $G_L^{(1)}(z)$ is the limit of the continued fraction: $G_L^{(1)}(z) = \lim_{M\rightarrow\infty}G_C^{(1)}(z,\theta^{(1)}; M)$. 

Using Mathematica we can show that for $M=2l-2$ with $l$ being a positive integer, 
\begin{align}
    \frac{P_{2l-2}}{Q_{2l-2}} = \frac{z^{2l+1}+2lAz^{2l+1}+A^2\sum_{i=1}^{2l} iz^i}{z^{2l+1}+A[-1+(z-1)\sum_{i=1}^{2l} iz^i]}. 
\end{align}
From Mathematica,
\begin{align}
    \sum_{i=1}^{2l} iz^i = \frac{z+[-1+2l(z-1)]z^{2l+1}}{(z-1)^2}.
\end{align}
Therefore, the truncated continued fraction is
\begin{align}
    &G_C^{(1)}(z,\theta^{(1)}; 2l-2)=\nonumber\\
    &\frac{zA^2 + z^{2l+1}(z-1+A)(-1-A-2Al+z+2Alz)}{(z-1)A+z^{2l+1}(z-1)(-1-A-2Al+z+2Alz)}.
\end{align}
By taking the large $l$ limit,
\begin{align}
    \lim_{l\rightarrow\infty} G_C^{(1)}(z,\theta^{(1)}; 2l-2) = \frac{z-1+A}{z-1} = G_L^{(1)}(z),
\end{align}
where $|z|>1$. The limit indeed converges to the discrete Laplace transformation result. Next, we check the cases for $M \in$ odd.
In general, for $M=2l-1$ with $l$ being positive integer, we find
\begin{align}
    \frac{P_{2l-1}}{Q_{2l-1}} = \frac{Az[(\sum_{i=0}^{2l-1}z^i)+A\sum_{i=1}^{2l-1}(2l-i)z^{i-1}]}{-1+A(-2l+\sum_{i=0}^{2l}z^i)}.
\end{align}
From Mathematica,
\begin{align}
    &\sum_{i=0}^{2l-1}z^i = \frac{z^{2l}-1}{z-1};\quad\quad
    \sum_{i=0}^{2l}z^i = \frac{z^{2l+1}-1}{z-1};\\
    &\sum_{i=1}^{2l-1}(2l-i)z^{i-1} = \frac{z^{2l}-2lz+2l-1}{(z-1)^2}.
\end{align}
Therefore, the truncated continued fraction is
\begin{align}
    &G_C^{(1)}(z,\theta^{(1)}; 2l-1) =\nonumber\\
    &\frac{Az[(-1+A-2Al+2Alz)+z-z^{2l}(z-1+A)]}{(z-1)[(-1+A-2Al+2Alz)+z-Az^{2l+1}]}.
\end{align}
By taking the large $l$ limit,
\begin{align}
    \lim_{l\rightarrow\infty} G_C^{(1)}(z,\theta^{(1)}; 2l-1) = \frac{z-1+A}{z-1} = G_L^{(1)}(z).
\end{align}
Therefore, the continued fraction $G_C^{(1)}(z,\theta^{(1)})$ is equivalent to $G_L^{(1)}(z)$.

\section{Reflection of continued fraction}
\label{Sec: C}
From \eqref{Eq: alpha beta-1} and \eqref{Eq: alpha beta-2}, one can show the following relations
\begin{subequations}
\label{Eq: alpha beta sym}
\begin{align}
    &\alpha_0(-z) = -\alpha_0(z);\\
    &\alpha_{k\geq 1}(-z,\theta_k) = \alpha_{k\geq 1}(z,\theta_k) = \alpha_{k\geq 1}(z,\pi-\theta_k)\\
    &\beta_0(-z,\theta_1) = -\beta_0(z,\pi-\theta_1);\\
    &\beta_{k\geq 1}(-z,\theta_k,\theta_{k+1}) \nonumber\\
    &=\Big\{\begin{array}{cc}
    -\beta_{k\geq 1} (z,\pi-\theta_k,\theta_{k+1}) & \text{if}\ k\ \text{is odd};\\
     -\beta_{k\geq 1} (z,\theta_k,\pi-\theta_{k+1}) & \text{else}.
    \end{array}
\end{align}
\end{subequations}
Note that $\theta_k \in [0,\pi]$. Since we already know that $G_L^{(2)}(z) = G_L^{(1)}(-z)$, this relation still holds in the continued fraction representation
\begin{align}
    &G_C^{(2)}(z,\theta^{(2)}) = G_C^{(1)}(-z,\theta^{(1)})\nonumber\\
    &= \frac{-\alpha_0(z)}{-\beta_0(z,\pi-\theta_1^{(1)}) + \frac{\alpha_1(z,\pi-\theta_1^{(1)})}{-\beta_1(z,\pi-\theta_1^{(1)},\theta_2^{(1)}) + \frac{\alpha_2(z,\theta_2^{(1)})}{ \ddots}}},
\end{align}
where we apply \eqref{Eq: alpha beta sym}. Then, we use the fact that the continued fraction is invariant under $\alpha_0 \rightarrow -\alpha_0$ and $\beta_k \rightarrow -\beta_k$ for all $k$
\begin{align}
    G_C^{(2)}(z,\theta^{(2)}) = \frac{\alpha_0(z)}{\beta_0(z,\pi-\theta_1^{(1)}) + \frac{\alpha_1(z,\pi-\theta_1^{(1)})}{\beta_1(z,\pi-\theta_1^{(1)},\theta_2^{(1)}) + \frac{\alpha_2(z,\theta_2^{(1)})}{ \ddots}}}.
\end{align}
Therefore, one concludes that
\begin{align}\label{zmz}
    \theta^{(2)}_k = \Big\{\begin{array}{cc}
    \pi - \theta^{(1)}_k & \text{if}\ k\ \text{is odd};\\
     \theta^{(1)}_k & \text{else}.
    \end{array}
\end{align}

\section{Square-reflection of continued fraction}
\label{Sec: D}
Since $G_L^{(4)}(z) = G_L^{(1)}(-z^2)$, it indicates that  $G_L^{(4)}(z)$ is invariant under reflection of $z$. In Appendix \ref{Sec: C}  it was shown that under $z \rightarrow -z$, the
angles transform as $\eqref{zmz}$. This implies, $\theta_k^{(4)} = \pi - \theta_k^{(4)}$ for $k \in$ odd. Therefore, $\theta_k^{(4)} = \pi/2$ for $k \in$ odd. Next, we consider $k \in $ even, $k=2l$. We now prove that $\theta_{2l}^{(4)} = \theta_l^{(1)}$. For this, we consider the truncated continued fraction, see definition in \eqref{Eq: truncated G},

\begin{subequations}
\begin{align}
    &G_C^{(4)}(z,\theta^{(4)};M) = \frac{\alpha_0^{(4)}}{\beta_0^{(4)} + \frac{\alpha_1^{(4)}}{  \ddots+\frac{\alpha_M^{(4)}}{\beta_M^{(4)}}}} = \frac{P_M^{(4)}}{Q_M^{(4)}};\\
    &G_C^{(1)}(-z^2, \theta^{(1)};M) = \frac{\alpha_0^{(1)}}{\beta_0^{(1)} + \frac{\alpha_1^{(1)}}{  \ddots+\frac{\alpha_M^{(1)}}{\beta_M^{(1)}}}} = \frac{P_M^{(1)}}{Q_M^{(1)}},
\end{align}
\end{subequations}
where the superscript in $\alpha, \beta, P$ and $Q$ is to emphasize the different underlying variable dependence. For example,
\begin{subequations}
\begin{align}
    &\alpha_0^{(4)}(z) = z;\\
    &\alpha_{k\geq 1}^{(4)}(z,\theta_k^{(4)}) = \Big\{\begin{array}{cc}
    1 & \text{if}\ k\ \text{is odd};\\
     z^2\sin^2\theta_k^{(4)} & \text{else}.
    \end{array}\\
    &\beta_0^{(4)}(z,\theta_1^{(4)}) = z;\\
    &\beta_{k\geq 1}^{(4)}(z,\theta_k^{(4)},\theta_{k+1}^{(4)}) = \Big\{\begin{array}{cc}
    z\cos\theta_{k+1}^{(4)} & \text{if}\ k\ \text{is odd};\\
     z\cos\theta_{k}^{(4)} & \text{else}.
    \end{array}\\
    &\alpha_0^{(1)}(-z^2) = -z^2;\\
    &\alpha^{(1)}_{k\geq 1}(-z^2,\theta_k^{(1)}) = \Big\{\begin{array}{cc}
    \sin^2\theta_k^{(1)} & \text{if}\ k\ \text{is odd};\\
     z^4\sin^2\theta_k^{(1)} & \text{else}.
    \end{array}\\
    &\beta_0^{(1)}(-z^2,\theta_1^{(1)}) = -z^2 - \cos\theta_1^{(1)};\\
    &\beta^{(1)}_{k\geq 1}(-z^2,\theta_k^{(1)},\theta_{k+1}^{(1)}) \nonumber\\
    &= \Big\{\begin{array}{cc}
    -z^2\cos\theta_{k+1}^{(1)} - \cos\theta_k^{(1)} & \text{if}\ k\ \text{is odd};\\
     -z^2\cos\theta_{k}^{(1)} - \cos\theta_{k+1}^{(1)} & \text{else}.
    \end{array}
\end{align}
\end{subequations}

Note that we have already used $\theta_{\text{odd}}^{(4)} =\pi/2$ in the above expressions for $\alpha^{(4)}$ and $\beta^{(4)}$. Let us conjecture that 
for any non-negative integer $M$
\begin{subequations}
\begin{align}
    &P_M^{(1)} = \Big\{\begin{array}{cc}
    P_{2M+1}^{(4)} & \text{if}\ M\ \text{is odd};\\
     -P_{2M+2}^{(4)}/z & \text{else}.
    \end{array},\\
    &Q_M^{(1)} = \Big\{\begin{array}{cc}
    Q_{2M+1}^{(4)} & \text{if}\ M\ \text{is odd};\\
     -Q_{2M+2}^{(4)}/z & \text{else}.
    \end{array},\label{Eq: square-reflection P Q}\\
   &G_C^{(1)} (-z^2,\theta^{(1)};M) \nonumber\\
   &= \Big\{\begin{array}{cc}
    G_C^{(4)}(z,\theta^{(4)},2M+1) & \text{if}\ M\ \text{is odd};\\
     G_C^{(4)}(z,\theta^{(4)},2M+2) & \text{else}.
    \end{array}
    \label{Eq: square-reflection G}
\end{align}
\end{subequations}
To prove this conjecture, we will use mathematical induction along with requiring $\theta_{2l}^{(4)} = \theta_l^{(1)}$. Since $Q$ obeys the same recursive equation as $P$, one needs only prove the above relation for $P^{(1)}$ and $P^{(4)}$, and the relation for $Q^{(1)}$ and $Q^{(4)}$ follows. 

First, we assume that the conjectured relation between $P^{(1)}$ and $P^{(4)}$ is true for $M=1$ up to some odd integer $M^*$. We are going to show that $P_{M^*+1}^{(1)} = -P_{2M^*+4}^{(4)}/z$ is also true. For $P_{M^*+1}^{(1)}$,
\begin{align}
    P_{M^*+1}^{(1)} &= \beta_{M^*+1}^{(1)}P_{M^*}^{(1)} + \alpha_{M^*+1}^{(1)}P_{M^*-1}^{(1)}\nonumber\\
    &= \beta_{M^*+1}^{(1)} P_{2M^*+1}^{(4)} -\frac{\alpha_{M^*+1}^{(1)}}{z}P_{2M^*}^{(4)}\nonumber\\
    &= (-z^2\cos\theta_{M^*+1}^{(1)}-\cos\theta_{M^*+2}^{(1)})P_{2M^*+1}^{(4)} \nonumber\\
    &\quad+ (-z^3\sin^2\theta_{M^*+1}^{(1)})P_{2M^*}^{(4)},
\end{align}
where we apply the recursive relation of $P^{(1)}$ in the first line and the assumption of induction in the second line. For $P_{2M^*+4}^{(4)}$, we apply the recursive relation down to $P_{2M^*+1}^{(4)}$ and $P_{2M^*}^{(4)}$
\begin{align}
    &P_{2M^*+4}^{(4)} = \beta_{2M^*+4}^{(4)} P_{2M^*+3}^{(4)} + \alpha_{2M^*+4}^{(4)} P_{2M^*+2}^{(4)}\nonumber \\
    &=\beta_{2M^*+4}^{(4)}(\beta_{2M^*+3}^{(4)} P_{2M^*+2}^{(4)} + \alpha_{2M^*+3}^{(4)} P_{2M^*+1}^{(4)}) \nonumber\\
    &\quad+
    \alpha_{2M^*+4}^{(4)}(\beta_{2M^*+2}^{(4)} P_{2M^*+1}^{(4)}+\alpha_{2M^*+2}^{(4)} P_{2M^*}^{(4)})\nonumber\\
    &\ \, \vdots\nonumber\\
    &=(\beta_{2M^*+4}^{(4)} \beta_{2M^*+3}^{(4)} \beta_{2M^*+2}^{(4)} +\beta_{2M^*+4}^{(4)} \alpha_{2M^*+3}^{(4)} \nonumber\\
    &\quad+\alpha_{2M^*+4}^{(4)} \beta_{2M^*+2}^{(4)})P_{2M^*+1}^{(4)}\nonumber\\
    &\quad + (\beta_{2M^*+4}^{(4)} \beta_{2M^*+3}^{(4)} \alpha_{2M^*+2}^{(4)} +\alpha_{2M^*+4}^{(4)} \alpha_{2M^*+2}^{(4)})P_{2M^*}^{(4)}\nonumber\\
    &= z(z^2\cos\theta_{2M^*+2}^{(4)}+\cos\theta_{2M^*+4}^{(4)})P_{2M^*+1}^{(4)} \nonumber\\
    &\quad+ (z^4\sin^2\theta_{2M^*+2}^{(4)})P_{2M^*}^{(4)}.
\end{align}
If $\theta_{2M^*+2}^{(4)} = \theta_{M^*+1}^{(1)}$ and $\theta_{2M^*+4}^{(4)} = \theta_{M^*+2}^{(1)}$, it is true that $P_{M^*+1}^{(1)} = -P_{2M^*+4}^{(4)}/z$.

Now we consider the case where $M^*$ is an even number. We are going to show that $P_{M^*+1}^{(1)} = P_{2M^*+3}^{(4)}$. $P_{M^*+1}^{(1)}$ obeys
\begin{align}
    &P_{M^*+1}^{(1)} = \beta_{M^*+1}^{(1)} P_{M^*}^{(1)} + \alpha_{M^*+1}^{(1)} P_{M^*-1}^{(1)}\nonumber\\
    &= -\frac{\beta_{M^*+1}^{(1)}}{z}P_{2M^*+2}^{(4)} +\alpha_{M^*+1}^{(1)} P_{2M^*-1}^{(4)}\nonumber\\
    &\ \, \vdots\nonumber\\
    &= -\frac{\beta_{M^*+1}^{(1)}}{z}(\beta_{2M^*+2}^{(4)} \beta_{2M^*+1}^{(4)}+\alpha_{2M^*+2}^{(4)})P_{2M^*}^{(4)} \nonumber\\
    &\quad + \left(-\frac{\beta_{M^*+1}^{(1)}}{z}\beta_{2M^*+2}^{(4)}\alpha_{2M^*+1}^{(4)} + \alpha_{M^*+1}^{(1)} \right) P_{2M^*-1}^{(4)}\nonumber\\
    &= z(\cos\theta_{M^*+1}^{(1)}+z^2\cos\theta_{M^*+2}^{(1)})P_{2M^*}^{(4)} \nonumber\\
    &\quad+ (\cos\theta_{M^*+1}^{(1)}\cos\theta_{2M^*+2}^{(4)} + \sin^2\theta_{M^*+1}^{(1)}\nonumber\\
    &\quad+z^2\cos\theta_{M^*+2}^{(1)} \cos\theta_{2M^*+2}^{(4)})P_{2M^*-1}^{(4)},
\end{align}
where we apply the recursive relation of $P^{(1)}$ in the first line and the assumption of induction in the second line. The remaining steps are to use the recursive relations of $P^{(4)}$.
For $P_{2M^*+3}^{(4)}$, we apply the recursive relation of $P^{(4)}$ down to $P_{2M^*}^{(4)}$ and $P_{2M^*-1}^{(4)}$
\begin{align}
    &P_{2M^*+3}^{(4)} \nonumber\\
    &= \beta_{2M^*+3}^{(4)} P_{2M^*+2}^{(4)} + \alpha_{2M^*+3}^{(4)} P_{2M^*+1}^{(4)}\nonumber\\
    &\ \, \vdots\nonumber\\
    &= (\beta_{2M^*+3}^{(4)} \beta_{2M^*+2}^{(4)} \beta_{2M^*+1}^{(4)} + \beta_{2M^*+3}^{(4)} \alpha_{2M^*+2}^{(4)} \nonumber\\
    &\quad+ \alpha_{2M^*+3}^{(4)} \beta_{2M^*+1}^{(4)})P_{2M^*}^{(4)}\nonumber\\
    &\quad + (\beta_{2M^*+3}^{(4)} \beta_{2M^*+2}^{(4)} \alpha_{2M^*+1}^{(4)} + \alpha_{2M^*+3}^{(4)} \alpha_{2M^*+1}^{(4)})P_{2M^*-1}^{(4)}\nonumber\\
    &= z(\cos\theta_{2M^*+2}^{(4)}+z^2\cos\theta_{2M^*+4}^{(4)})P_{2M^*}^{(4)} \nonumber\\
    &\quad+ (1+z^2\cos\theta_{2M^*+2}^{(4)}\cos\theta_{2M^*+4}^{(4)})P_{2M^*-1}^{(4)}.
\end{align}
If $\theta_{2M^*+2}^{(4)} = \theta_{M^*+1}^{(1)}$ and $\theta_{2M^*+4}^{(4)} = \theta_{M^*+2}^{(1)}$, we find that  $P_{M^*+1}^{(1)} = P_{2M^*+3}^{(4)}$.

By mathematical induction, we prove that \eqref{Eq: square-reflection P Q} and \eqref{Eq: square-reflection G} hold. As $M$ goes to infinity, we have the following relation for the continued fraction
\begin{align}
    G_C^{(4)}(z,\theta^{(4)}) = G_C^{(1)}(-z^2, \theta^{(1)}),
\end{align}
where  $\theta_{2l-1}^{(4)} = \pi/2$ and $\theta_{2l}^{(4)} = \theta_l^{(1)}$ for all positive integer $l$.

\section{Krylov angles \eqref{Eq: m=3 Krylov angle} of 3-period autocorrelation}\label{Sec: E}

Similar to the discussion in 
Appendix \ref{Sec: B}, \eqref{Eq: m=3 Krylov angle} is conjectured from solving the first few Krylov angles by equating $A^{(3)}(n) = (1|K^n|1)$. For $n = 1$,
\begin{align}
    -\frac{A}{2} = \cos\theta_1^{(3)}.
\end{align}
Therefore, one obtains $\cos\theta_1^{(3)} = -A/2$ and $\sin\theta_1^{(3)} = \sqrt{4-A^2}/2$. For $n =2$,
\begin{align}
    -\frac{A}{2} &= \cos^2\theta_1^{(3)} - \sin^2\theta_1^{(3)} \times \cos\theta_2^{(3)} \nonumber\\
    &= \frac{A^2}{4} - \frac{4-A^2}{4}\cos\theta_2^{(3)}.
\end{align}
The solutions are $\cos\theta_2^{(3)} = -A/(-2+A)$ and $\sin\theta_2^{(3)} = 2\sqrt{1-A}/(2-A)$. For larger $n$, we obtain solutions by symbolic computation on Mathematica and conjecture \eqref{Eq: m=3 Krylov angle} from the pattern of the solutions. Below, we show that the continued fraction with Krylov angle obeying \eqref{Eq: m=3 Krylov angle} converges to the discrete Laplace transformation
\begin{align}
    G_L^{(3)}(z) = \frac{1-A + (1-\frac{A}{2})z+z^2}{1+z+z^2}.
\end{align}
To show that $G_L^{(3)}(z) = \lim_{M\rightarrow\infty}G_C^{(3)}(z,\theta^{(3)}; M)$, we consider six different cases  $M=6l-6, 6l-5, 6l-4, 6l-3, 6l-2, 6l-1$ with non-negative integer $l$. For each case, we show the convergence to $G_L^{(3)}(z)$ as $l \to \infty$.
For $M = 6l-6$ and for $l\geq 2$, we have
\begin{widetext}
\begin{align}
    \frac{P_{6l-6}}{Q_{6l-6}} &= z\left[8z^{6l-6} + (48l-56)Az^{6l-6} \right. \nonumber\\
    &+ A^3\left(\sum_{i=1}^{2l-2}[6l-8+(33+36(l-2))(i-1)]z^{3i-3} -(6l-5)(3i-2)z^{3i-2} - (18l-24)iz^{3i-1}  \right) \nonumber\\
    &\left.\left. 2A^2 \left(\left(\sum_{i=1}^{2l-2} (6i-5)z^{3i-3} - (3i-2)z^{3i-2} -3iz^{3i-1}  \right) + 12(3l-4)(l-1)z^{6l-6} \right) \right]\right/ \nonumber\\
    &\left[ 8z^{6l-5} + 4A \left( 1 -\left(\sum_{i=1}^{2l-2} 2z^{3i-2} - z^{3i-1} -z^{3i} \right) + (12l-14)z^{6l-5} \right) \right. \nonumber\\
    &\left. 2A^2\left(6l-8 -\left(\sum_{i=1}^{2l-2} (12l-13)z^{3i-2} -(6l-5)z^{3i-1}-(6l-8)z^{3i} \right) + 12(3l-4)(l-1)z^{6l-5}  \right) \right].
\end{align}
From Mathematica,
\begin{align}
 \lim_{M\rightarrow\infty}G_C^{(3)}(z,\theta^{(3)}; 6l-6)&= \lim_{l\rightarrow\infty} \frac{P_{6l-6}}{Q_{6l-6}} = \frac{1-A + (1-\frac{A}{2})z+z^2}{1+z+z^2} = G_L^{(3)}(z).
\end{align}

For $M = 6l-5$ we have
\begin{align}
    &\frac{P_{6l-5}}{Q_{6l-5}}\nonumber\\ 
    &= Az\left[A\left((6l-7)+\left(\sum_{i=0}^{2l-3} (3i-1)z^{6l-6-3i} - (2+6i)z^{6l-7-3i}+(3i+3)z^{6l-8-3i}\right)\right) \right.\nonumber\\
    &\left.\left. + 2\left(1+z-\left(\sum_{i=1}^{2l-2}2z^{3i-1} -z^{3i}-z^{3i+1}\right)\right) \right]\right/ \left[4 + 2A\left((6l-5)-\left(\sum_{i=0}^{2l-2} 2z^{3i} -z^{3i+1}-z^{3i+2}\right)\right)\right].
\end{align}
From Mathematica,
\begin{align}
\lim_{M\rightarrow\infty}G_C^{(3)}(z,\theta^{(3)}; 6l-5)&= \lim_{l\rightarrow\infty} \frac{P_{6l-5}}{Q_{6l-5}} = \frac{1-A + (1-\frac{A}{2})z+z^2}{1+z+z^2} = G_L^{(3)}(z).
\end{align}
For $M = 6l-4$
\begin{align}
    \frac{P_{6l-4}}{Q_{6l-4}} = \frac{4z^{6l-3}+(12l-10)Az^{6l-3}-A^2z\left[ 2+z-(\sum_{i=1}^{2l-2}6iz^{3i-1} - (3i+2)z^{3i} -(3i+1)z^{3i+1}) \right]}{4z^{6l-3}+2A\left[ -2 + (\sum_{i=1}^{2l-1} z^{3i-2} + z^{3i-1} -2z^{3i}) + (6l-3)z^{6l-3} \right]}.
\end{align}
From Mathematica,
\begin{align}
   \lim_{M\rightarrow\infty}G_C^{(3)}(z,\theta^{(3)}; 6l-4)&= \lim_{l\rightarrow\infty} \frac{P_{6l-4}}{Q_{6l-4}} = \frac{1-A + (1-\frac{A}{2})z+z^2}{1+z+z^2} = G_L^{(3)}(z).
\end{align}
For $M = 6l-3$ 
\begin{align}
    \frac{P_{6l-3}}{Q_{6l-3}} &= Az\left[ A^2\left(\sum_{i=1}^{2l-1} 3i(6l-5)z^{6l-3-3i} + (3i-2)(6l-2)z^{6l-2-3i} - (6l-5+(36l-21)(i-1))z^{6l-1-3i}\right) \right.\nonumber\\
    & +2A\left(12l-8+\sum_{i=1}^{2l-1}(6l-4+3i)z^{6l-2-3i} -6(2l-2 + i)z^{6l-1-3i} + (6l-8+3i)z^{6l-3i}\right) \nonumber\\
    &\left.\left.+4 \left( 1+ \sum_{i=1}^{2l-1} z^{3i-2} -2z^{3i-1}+z^{3i} \right) \right]\right/ \nonumber\\
    &\left[8 + 4A \left( 12l-8+z + \sum_{i=1}^{2l-1} z^{3i-1} -2z^{3i} + z^{3i+1} \right) \right.\nonumber\\
    &\left.+ 2A^2\left( 15-48 l+36 l^2 + (6l-5)z + \sum_{i=1}^{2l-1}(6l-2)z^{3i-1} - (12l-7)z^{3i} +(6l-5)z^{3i+1}\right) \right].
\end{align}
From Mathematica,
\begin{align}
\lim_{M\rightarrow\infty}G_C^{(3)}(z,\theta^{(3)}; 6l-3)&= \lim_{l\rightarrow\infty} \frac{P_{6l-3}}{Q_{6l-3}} = \frac{1-A + (1-\frac{A}{2})z+z^2}{1+z+z^2} = G_L^{(3)}(z).
\end{align}
For $M = 6l-2$ 
\begin{align}
    \frac{P_{6l-2}}{Q_{6l-2}} = \frac{4z^{6l-1}+(12l-8)Az^{6l-1}+A^2z\left[1 + (\sum_{i=1}^{2l-1}(6i-4)z^{3i-2} -3i z^{3i-1} -(3i-1)z^{3i})  \right]}{4z^{6l-1}+2A[(\sum_{i=0}^{2l-1}  z^{3i} +z^{3i+1} -2z^{3i+2}) + (6l-2)z^{6l-1}]}
\end{align}
From Mathematica,
\begin{align}    \lim_{M\rightarrow\infty}G_C^{(3)}(z,\theta^{(3)}; 6l-2)&= \lim_{l\rightarrow\infty} \frac{P_{6l-2}}{Q_{6l-2}} = \frac{1-A + (1-\frac{A}{2})z+z^2}{1+z+z^2} = G_L^{(3)}(z).
\end{align}
For $M = 6l-1$
\begin{align}
    \frac{P_{6l-1}}{Q_{6l-1}} = \frac{Az(1+2z)\left[A \left(6l-2 - \left(\sum_{i=1}^{2l-1} 3iz^{6l-3i-2} -(3i-2)z^{6l-3i}  \right)\right) +2 \left(\sum_{i=1}^{2l}z^{3i-3} -z^{3i-2}\right)\right]}{4 + A\left[12l-4 +\left( \sum_{i=1}^{2l} 2z^{3i-2} + 2z^{3i-1} -4z^{3i} \right) \right]}.
\end{align}
This gives
\begin{align}    \lim_{M\rightarrow\infty}G_C^{(3)}(z,\theta^{(3)}; 6l-1)&= \lim_{l\rightarrow\infty} \frac{P_{6l-1}}{Q_{6l-1}} = \frac{1-A + (1-\frac{A}{2})z+z^2}{1+z+z^2} = G_L^{(3)}(z).
\end{align}
Therefore, the Krylov angle \eqref{Eq: m=3 Krylov angle} reproduce the Laplace transformation $G_L^{(3)}(z)$.
\end{widetext}
 

%

\end{document}